\documentclass{emulateapj}
\shorttitle{Influence of irradiation-driven winds on the evolution of IMBH X-ray binaries}
\shortauthors{Han, Jiang \& Chen}

\begin{document}


\title{Influence of irradiation-driven winds on the evolution of intermediate-mass black hole X-ray binaries }


\author{Xiao-Qin Han$^{1}$, Long Jiang$^{2,3,1}$, and Wen-Cong Chen$^{2,1}$}
\affil{$^1$ School of Physics and Electrical Information, Shangqiu Normal University, Shangqiu 476000, China;\\
$^2$ School of Science, Qingdao University of Technology, Qingdao 266525, China;
chenwc@pku.edu.cn\\
$^3$ Xinjiang Astronomical Observatory, CAS, Urumqi, Xinjiang 830011, China.
}



\begin{abstract}
In young dense clusters, an intermediate-mass black hole (IMBH) may get a companion star via exchange encounters or tidal capture, and then evolves toward IMBH X-ray binary by the Roche lobe overflow. It is generally thought that IMBH X-ray binaries
are potential ultra-luminous X-ray sources (ULXs), hence their evolution is very significant. However, the irradiation-driven winds by the strong X-ray flux from the accretion disks around the IMBHs play an important role in determining the evolution of IMBH X-ray binaries, and should be considered in the detailed binary evolution simulation. Employing the models with the MESA code, we focus on the influence of irradiation-driven winds on the evolution of IMBH X-ray binaries. Our simulations indicate that a high wind-driving efficiency ($f=0.01$ for $Z=0.02$, and $f=0.002$ for $Z=0.001$) substantially shorten the duration in the ULX stage of IMBH X-ray binaries with an intermediate-mass ($5~M_{\odot}$) donor star. However, this effect can be ignored for high-mass ($10~M_{\odot}$) donor stars. The irradiation effect ($f=0.01$ or $0.002$) markedly shrink the initial parameter space of IMBH binaries evolving toward ULXs with high luminosity ($L_{\rm X}>10^{40}~\rm erg\,s^{-1}$) and hyperluminous X-ray sources in the donor-star mass versus orbital period diagram. Furthermore, the irradiation effect results in an efficient angular momentum loss, yielding to IMBH X-ray binaries with relatively close orbits. In our simulated parameter space, about 1\% of IMBH binaries would evolve toward compact X-ray sources owing to short initial orbital periods, some of which might be detected as low-frequency gravitational wave sources.
\end{abstract}


\section{Introduction}
Stellar-mass black holes (BHs,$\sim10~ M_{\odot}$) are the evolutionary products of massive stars that have exhausted all their nuclear fuel. The first gravitational-wave (GW) event GW150914 discovered by the advanced Laser Interferometer Gravitational-Wave Observatory (aLIGO) \citep{abbo16a,abbo16b} revealed that stellar-mass ($\la100~ M_{\odot}$) BHs widely exist in the universe. Meanwhile, many observations have revealed supermassive BHs ($\ga10^{6}~ M_{\odot}$) at the centers of most galaxies \citep{korm95}. A continuous BH mass spectrum would suggest the existence of a transition population with masses in the range of $100~M_{\odot}\la M_{\rm BH}\la10^{6}~ M_{\odot}$, which is referred to as intermediate-mass black holes (IMBHs). Recently, the analysis of dynamical states of the globular cluster 47 Tucanae and NGC 6624 probed using pulsars provided evidences of the existence of IMBHs \citep{kizi17,pere17}.

The concept of IMBHs is tightly related to the ultra-luminous X-ray sources (ULXs). \emph{ROSAT, Chandra}, and \emph{XMM-Newton} observed many ULXs with X-ray luminosities in the range of $10^{39}$ to $10^{41}~\rm erg\,s^{-1}$, which are off-nucleus X-ray sources in external galaxies \citep{fabb89,robe00,ptak04,fabb06}. The Eddington luminosity of a stellar-mass BH accretor with a mass of $10~ M_{\odot}$ is $ 1.5-3.0\times10^{39}~\rm erg\,s^{-1}$, depending on the chemical composition of the accreting material. Considering that the Eddington luminosity of an IMBH with a mass of $1000~ M_{\odot}$ is $\sim 10^{41}~\rm erg\,s^{-1}$, \cite{colb99} firstly proposed that the accretors in some high-luminosity ULXs are IMBHs. Furthermore, the cool thermal radiation signatures provided by the X-ray spectra in some ULXs are consistent with low inner-disk temperatures in the accretion-disk models of IMBHs \citep{kaar03,mill03,mill04,crop04,feng11}. In particular, observations for hyperluminous X-ray sources (HLXs, $L_{\rm X}>0^{41}~\rm erg\,s^{-1}$) presented sound evidences for the existence of IMBHs. Several reliable HLXs are M82 X-1 \citep{mats01}, S0/a galaxy ESO 243-49 HLX-1 \citep{farr09,wier10,davi11,serv11,sutt12}, Cartwheel N10 \citep{pizz10}, and CXO J122518.6+144545 \citep{jonk10}.

\cite{port04b} suggested that HLX M82 X-1 is a $\sim1000~ M_{\odot}$ IMBH accreting from a $10-15~M_{\odot}$ donor star near the end of its main sequence or slightly evolved state. However, \cite{patr06} found that a $\sim200-5000~ M_{\odot}$ IMBH accreting material from a $20-25~M_{\odot}$ giant companion via Roche-lobe overflow can explain the observed phenomenon of HLX M82 X-1. HLX-1 near the spiral galaxy ESO 243-49 is currently thought to be the best candidate for an IMBH. BH accretion disk models have been used to fit the X-ray spectrum of HLX-1 \citep{davi11,serv11,gode12,stra14}, where the IMBH mass was estimated to be $\sim10^{4}-10^{5}~ M_{\odot}$ \citep[see also][]{webb12}. On the basis of the optical spectroscopic observations, \cite{sori13} proposed that HLX-1 and its surrounding stars originated from either a disrupted dwarf galaxy or a nuclear recoil.

Physical collisions in young dense star clusters are frequent when the stellar density in the core is high \citep{quin87,port99}. In this situation, it is possible for an IMBH to form through runaway collision in $3-5$ Myr \citep{port02,port04a,saku17}. In a young ($\la10~\rm Myr$) star cluster with extreme density ($\rho\ga10^{5}~M_{\odot}\rm pc^{-3}$) and many stars ($N>10^{5}$), tidal or dynamical capture readily occurs \citep{port04a,baum06}. As a result of tidal capture or exchange encounters, the main sequence stars can spiral into and circularize around an IMBH \citep{hopm04}. As an example,
the ULX source in the young cluster MGG-11 of starburst galaxy M82 is likely to be an IMBH accreting from a captured donor star, and $\ga10\%$ of similar clusters with IMBHs acquire a tidally captured star that had already been circularized \citep{hopm04}.

Through the evolutionary simulation of IMBH X-ray binaries formed in dense star clusters via tidal capture, \cite{li04} found that IMBHs can appear as ULXs with luminosities exceeding $10^{40}~\rm erg\,s^{-1}$ over several Myr. \cite{madh06} calculated 30,000 binary evolution models of IMBH X-ray binaries, and obtained two requirements for the donor stars captured by IMBHs to produce active ULXs with luminosities $\ga 10^{40}~\rm erg\,s^{-1}$. First, the donor stars of the IMBHs must have a mass $\ga 8~M_{\odot}$. Second, the initial orbital separations of IMBH binaries after circularization should be about 6-30 times the radii of the donor stars when they are on the zero-age main-sequence (ZAMS). N-body simulations showed that the tidal energy dissipation of stars in young star clusters can result in IMBH binaries like those mentioned above \citep{baum06}.

In principle, the strong X-ray flux from the accretion disks around IMBHs should irradiate the surfaces of the donor stars, and generate strong stellar winds \citep[see also][]{rude89,tava93}. This irradiation effect can not be ignored because of the high X-ray luminosities of IMBH X-ray binaries. In this work, we focus on the effect of irradiation-driven winds on the evolution of IMBH X-ray binaries, and explore the initial parameter space of IMBH binaries (consisting of an IMBH and a companion star) that can evolve toward the observed ULXs.

\section{Description of IMBH binary evolution}
\subsection{Binary evolution code}
In this work, we simulate the evolution of IMBH X-ray binaries using
a MESAbinary update version (r10108) in the Modules for Experiments in Stellar
Astrophysics code \citep[MESA;][]{paxt11,paxt13,paxt15}. As an evolutionary beginning,
IMBH binaries consisting of an IMBH (with a mass of
$M_{\rm bh}$) and a main sequence companion star (with a mass of $M_{\rm d}$)
are thought to have already formed in circular orbits via exchange encounters or tidal capture in dense clusters.
The IMBH is assumed to be a point mass with a constant mass of $1000~M_{\odot}$. We consider two metallicities of the donor stars, that is $Z=0.02$, and 0.001. The effective
Roche-lobe radius of the companion star satisfies \citep{egg83}
\begin{equation}
\frac{R_{\rm L}}{a}=\frac{0.49q^{2/3}}{0.6q^{2/3}+ {\rm
ln}(1+q^{1/3})},
\end{equation}
where $a$ is the orbital separation, and $q = M_{\rm d}/M_{\rm bh}$ is the
mass ratio of the IMBH binary. If the donor star fills its Roche lobe, the mass transfer rate
is determined according to the prescription given by \cite{ritt88}. We execute additional models (see subsections 2.2, 2.3, and 2.4) using the files run$\_$binary$\_$extras.f90 and run$\_$star$\_$extras.f90 (We also share these two files with the MESA community).

We perform many detailed binary evolution simulations of IMBH binaries. A donor star would be tidally disrupted when the separation to the IMBH is less than the tidal radius $R_{\rm t}=(M_{\rm bh}/M_{\rm d})^{1/3}R_{\rm d}$. According to the law of the conservation of orbital angular momentum, the circularization radius is $a_{\rm cir}\approx(4-5) R_{\rm t}$ after in-spiraling and circulation of the donor star \citep{hopm04,li04}. Taking $M_{\rm d}=1.0~M_{\odot}$, we have $a_{\rm cir}\approx(40-50) R_{\odot}$, which corresponds to a minimum orbital period of $0.9-1.3$ days. \cite{blec06} obtained a wide range of separations ($100-10^{4}~R_{\odot}$) between captured donor stars and IMBHs; this range gives a maximum orbital period of several thousand days. Therefore, the donor-star masses and initial orbital periods are assumed to be $M_{\rm d,i}=1-20~M_{\odot}$ and ${\rm log}(P_{\rm orb, i}/\rm days)=0-3.0$ in the simulations.

\subsection{Isotropic winds}
The donor star fills its Roche lobe as it thermally expands of the donor star during nuclear burning or angular momentum is lost from the binary system. The H-rich material on the surface of the donor star
is transferred onto the IMBH at a rate of $\dot{M}_{\rm tr}$ through the inner Lagrange point. During the mass transfer,
the maximum accretion rate of the IMBH is limited to the Eddington rate
\begin{equation}
\dot{M}_{\rm Edd}=2.6\times10^{-5}\frac{M_{\rm bh}}{1000 M_{\odot}}\left(\frac{0.1}{\eta}\right)\left(\frac{1.7}{1+X}\right) M_{\odot}\,\rm yr^{-1},
\end{equation}
where $X$ is the H abundance in the outer layers of the donor star. Because the mass growth of the IMBH can be ignored,
the energy conversion efficiency $\eta=1-\sqrt{1-(M_{\rm bh}/3M_{\rm bh,0})^{2}}\approx0.06$ \citep{pods03}, here $M_{\rm bh,0}$
is the initial IMBH mass.

If the mass transfer rate is greater than the Eddington rate, the transferred matter that exceeds the Eddington rate
is assumed to be ejected in the vicinity of the IMBH. This materials forms isotropic winds, and carries away the specific
orbital-angular-momentum of the IMBH \citep{taur99}.

\subsection{X-ray luminosity}
There exist a high/soft state (with a soft and thermally-dominated spectrum) and low/hard state (with a non-thermal hard power
law spectrum) in the observed spectrum of BH X-ray binaries \citep{nowa95}.
The transition between the low and high states should relate to a critical accretion rate $\dot{M}_{\rm crit}$ \citep{kord02}.
If the accretion rate $\dot{M}_{\rm acc}>\dot{M}_{\rm crit}$, the disk luminosity is directly proportional to $\dot{M}_{\rm acc}$
like for a standard accretion disk; in contrast, the disk luminosity is directly proportional to $\dot{M}_{\rm acc}^{2}$
for the expected optically thin, advection-dominated accretion flow \citep{nara95}. The X-ray luminosity of the accretion disk can
thus be written as \citep{kord02}
\begin{equation}
L_{\rm X}= \left\{\begin{array}{l@{\quad}l}\epsilon\dot{M}_{\rm acc}c^{2},
& \dot{M}_{\rm crit}<\dot{M}_{\rm acc}\leq \dot{M}_{\rm Edd} \strut\\
\epsilon\left(\frac{\dot{M}_{\rm acc}}{\dot{M}_{\rm crit}}\right)\dot{M}_{\rm acc}c^{2}
, & \dot{M}_{\rm acc}<\dot{M}_{\rm crit}, \strut\\\end{array}\right.
\end{equation}
where $\epsilon$ is the radiative efficiency of the standard accretion disk. In this work, we
take a typical critical accretion rate $\dot{M}_{\rm crit}=10^{-7}~M_{\odot}\,\rm yr^{-1}$ \citep{nara95,hopm04}, and $\epsilon=0.1$.

\begin{figure*}
\centering
\begin{tabular}{cc}
\includegraphics[width=0.5\textwidth,trim={10 10 30 30},clip]{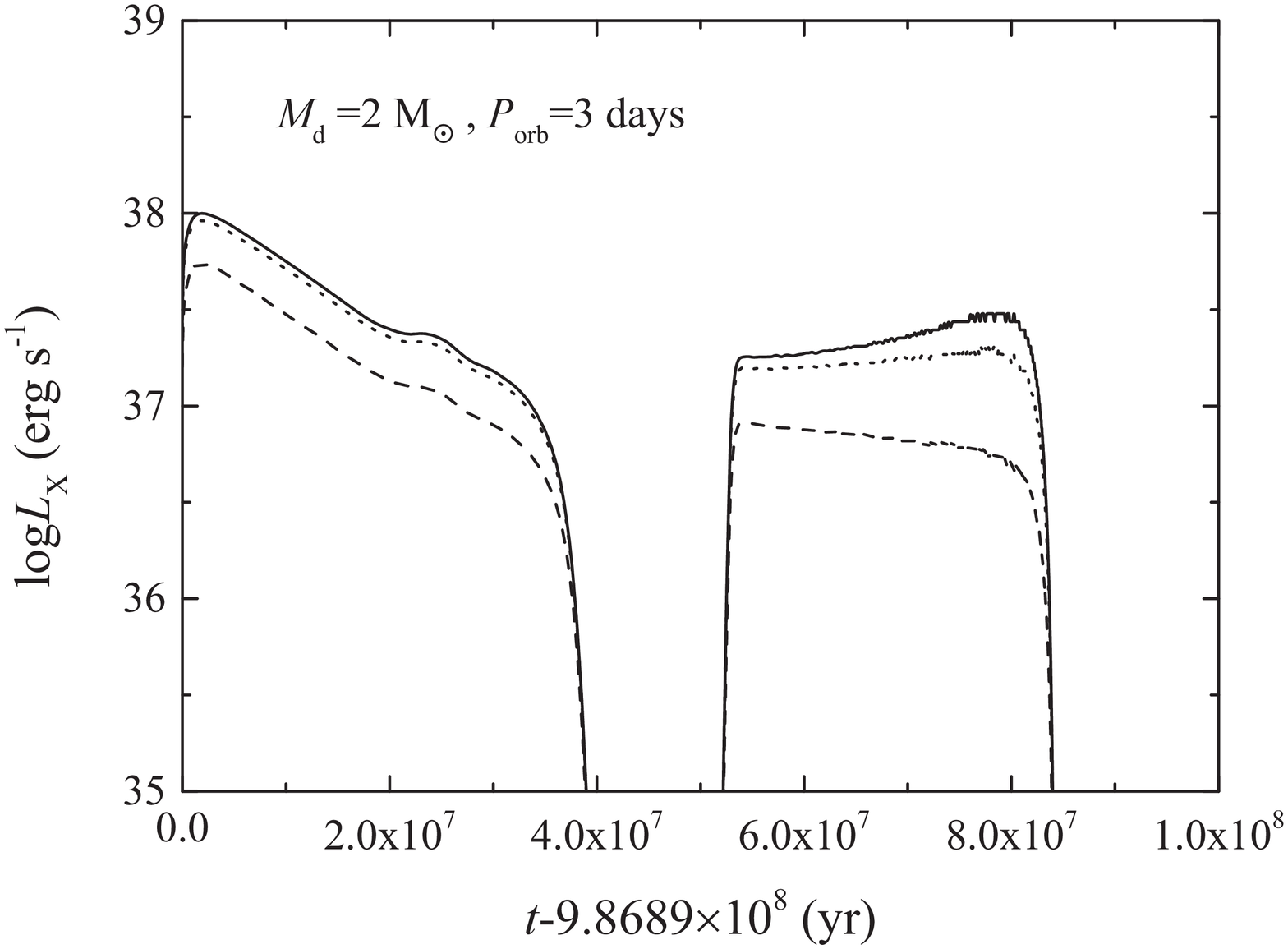} &   \includegraphics[width=0.5\textwidth,trim={10 10 30 30},clip]{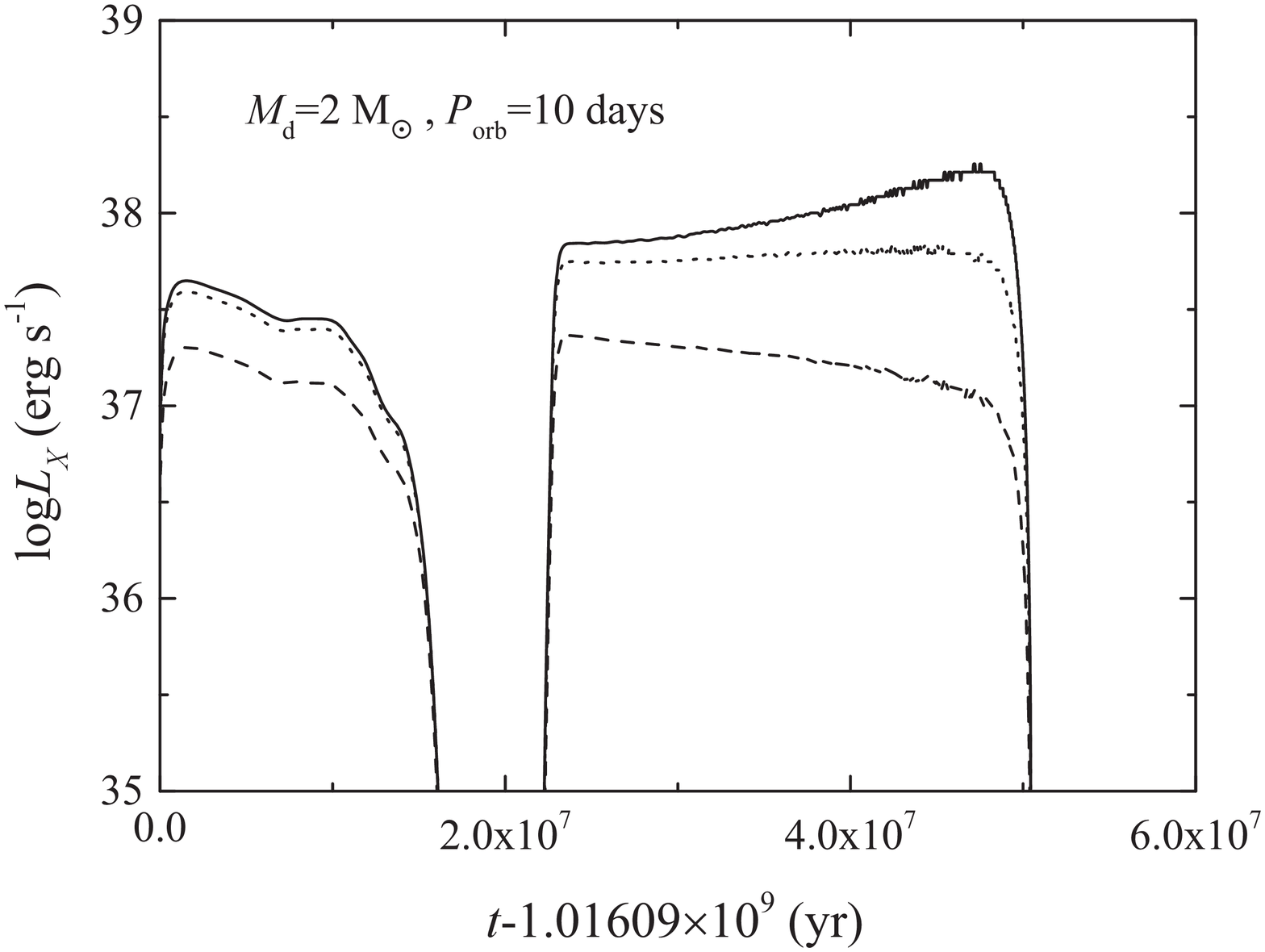} \\
\includegraphics[width=0.5\textwidth,trim={10 10 30 30},clip]{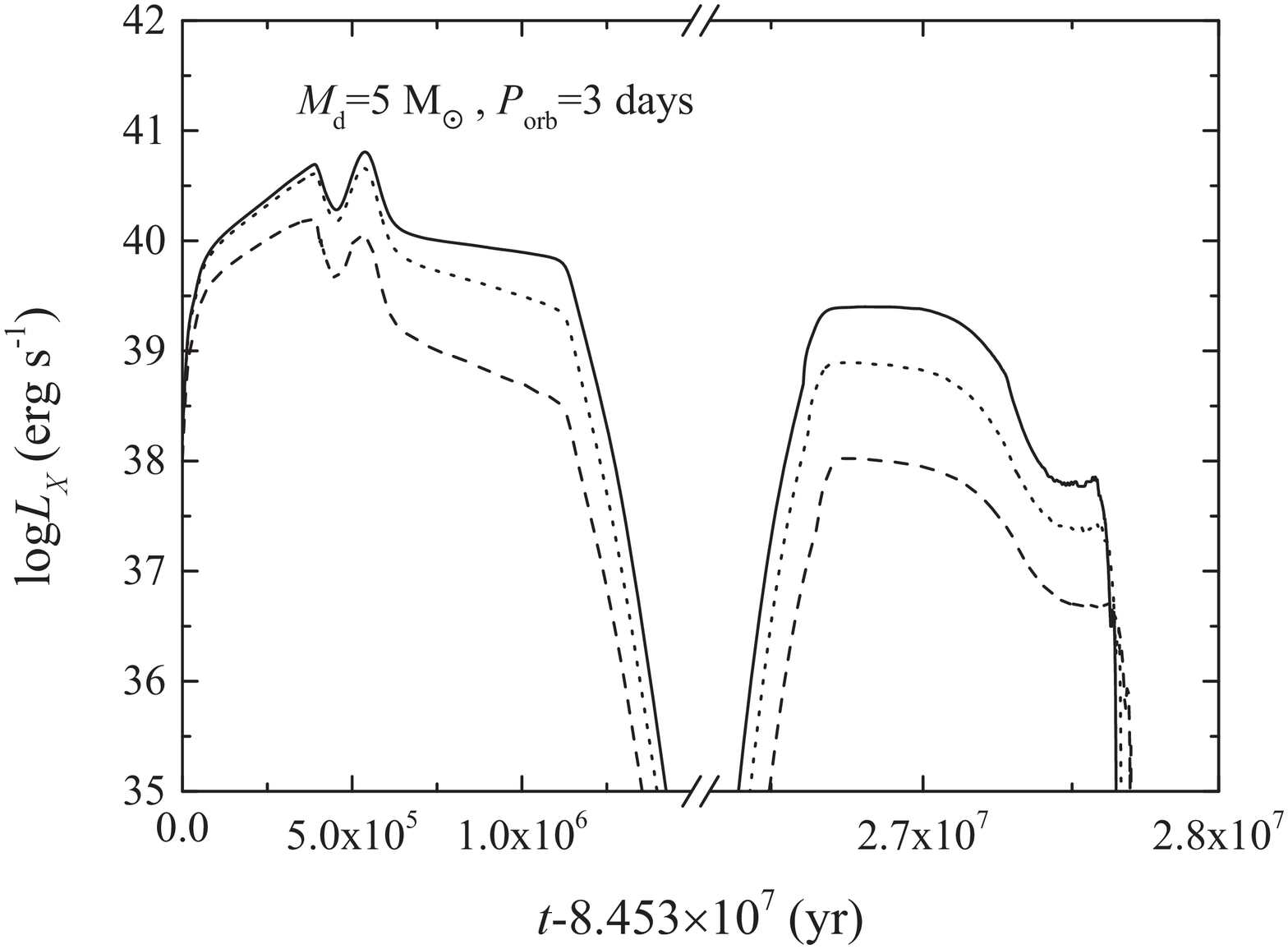} &   \includegraphics[width=0.5\textwidth,trim={10 10 30 30},clip]{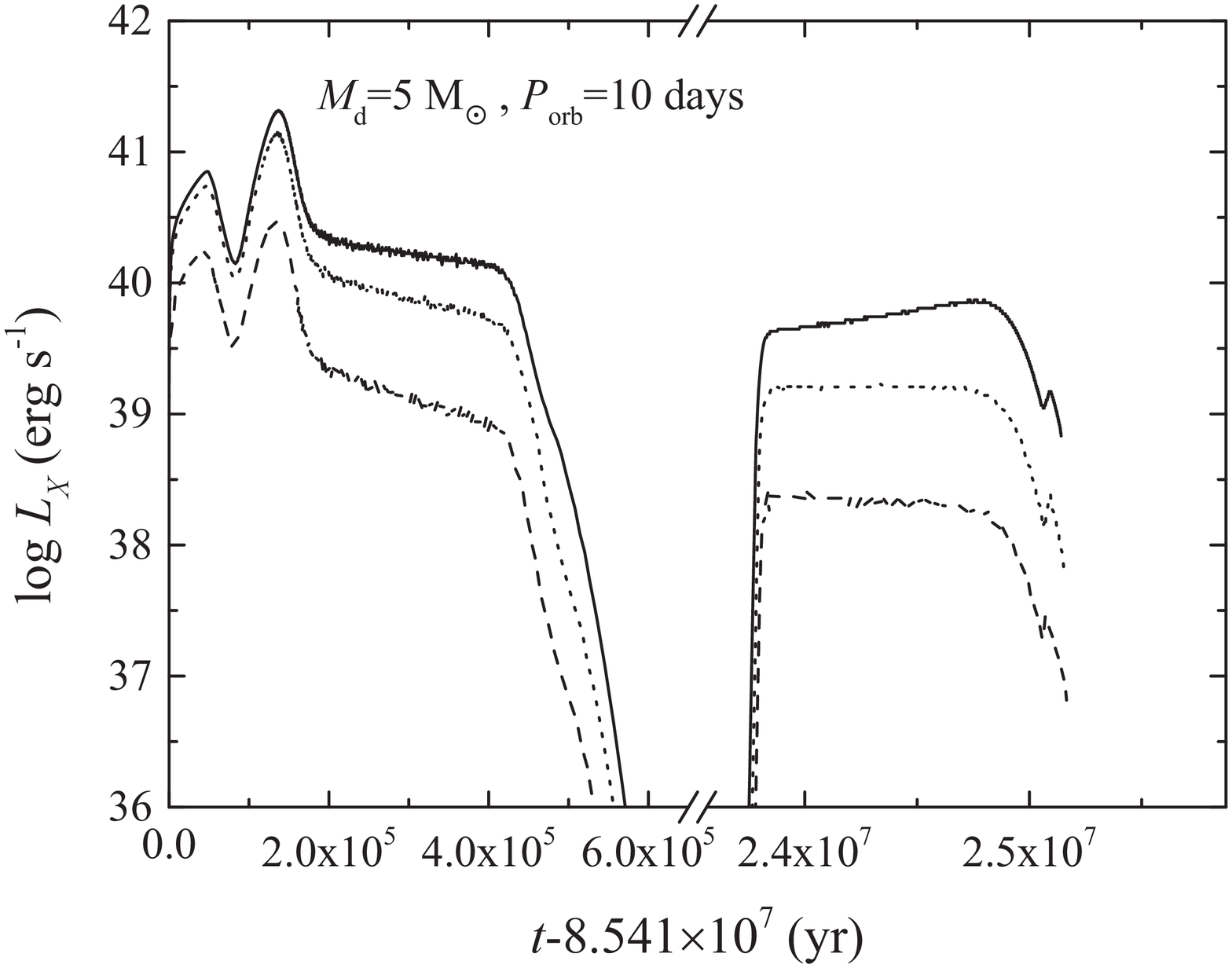} \\
\includegraphics[width=0.5\textwidth,trim={10 10 30 30},clip]{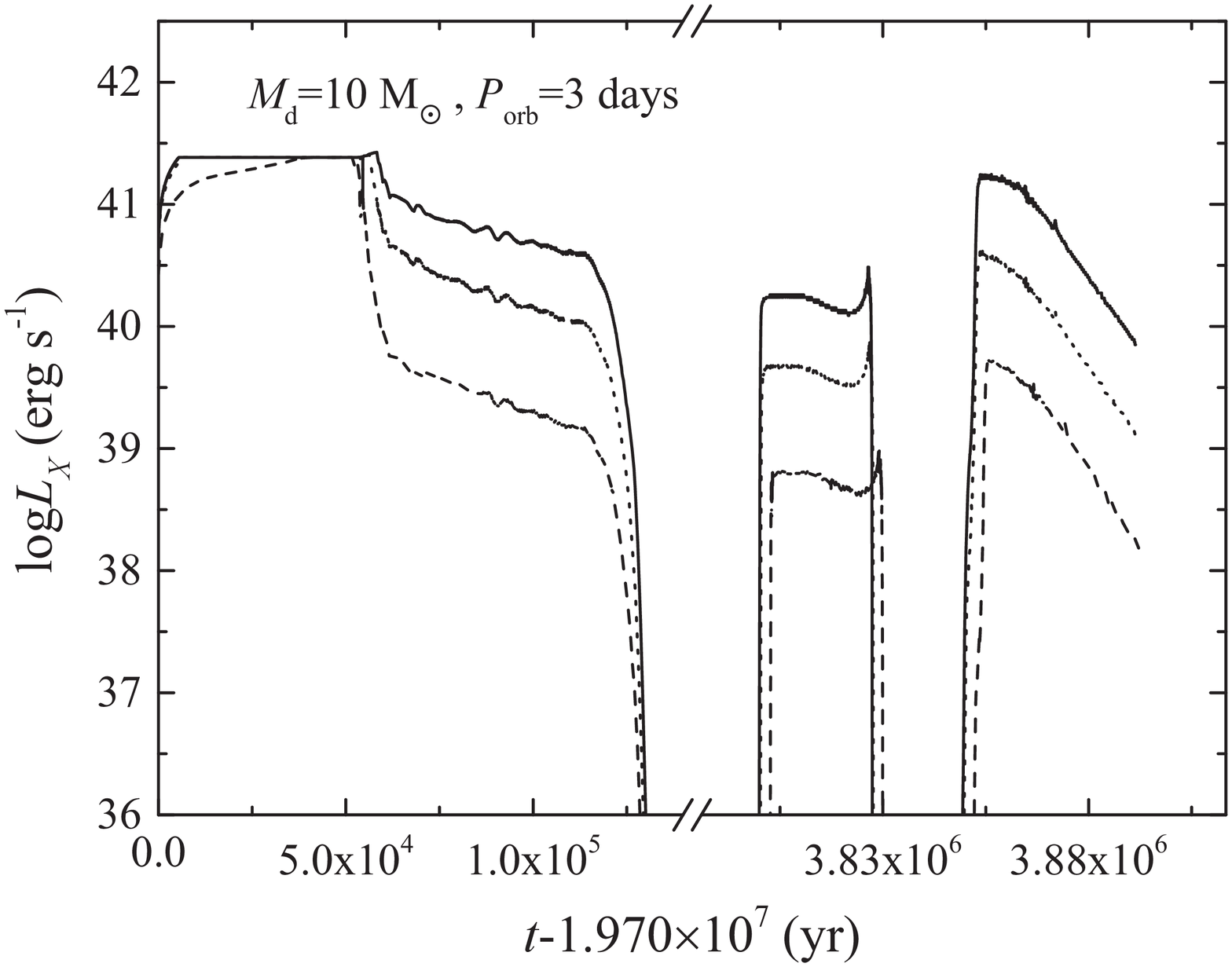} &   \includegraphics[width=0.5\textwidth,trim={10 10 30 30},clip]{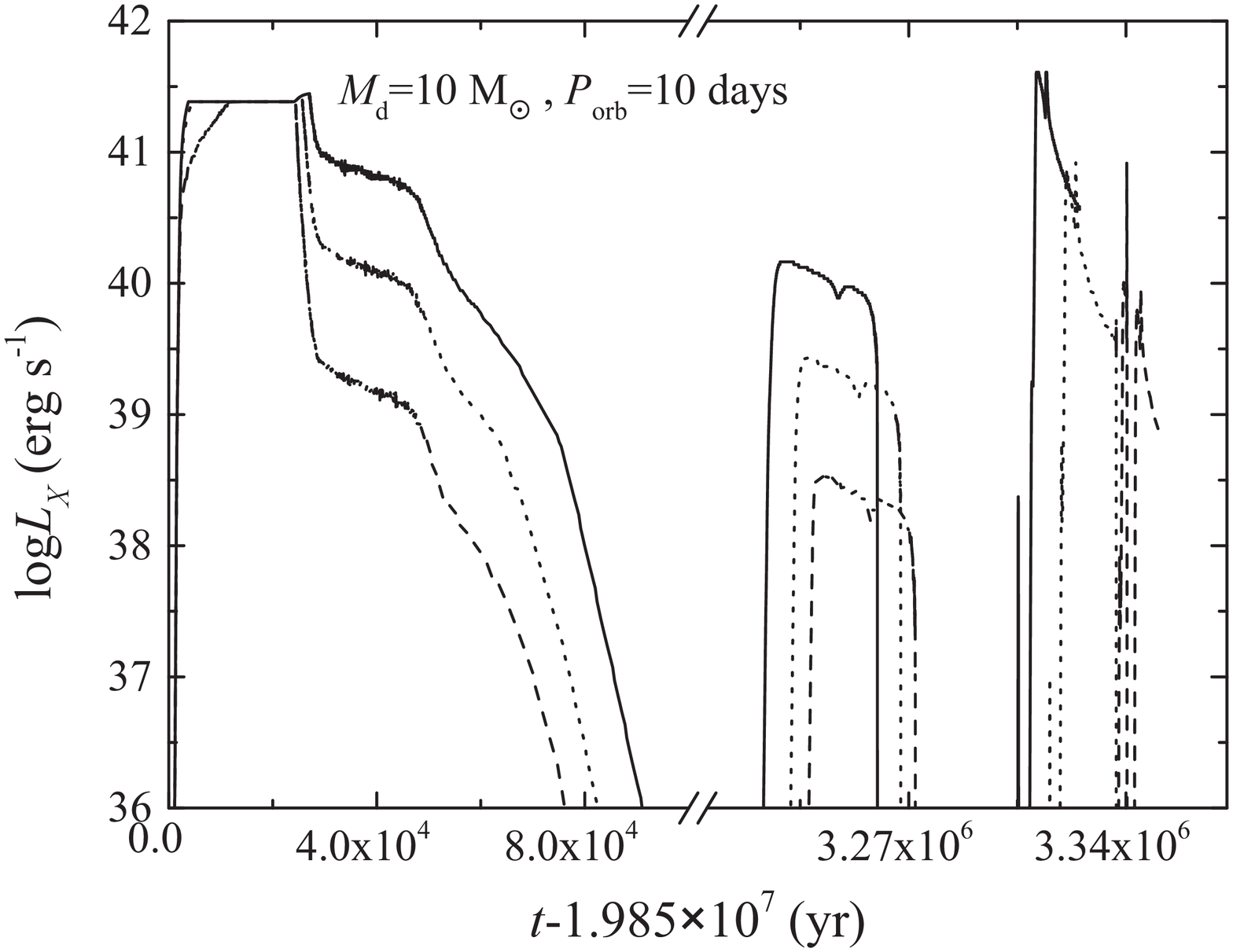} \\
\end{tabular}
\caption{\label{fig:pulses}Evolution of X-ray luminosity as a function of the mass transfer timescale
for IMBH binaries with different initial donor-star ($Z=0.02$) masses
and initial orbital periods. The solid, dashed, and dotted curves denote the wind-driving
efficiency $f=0, 0.01$, and 0.001, respectively. }
\end{figure*}

\subsection{Irradiation-driven winds}
In a compact binary, the stellar wind from the donor star generated by X-ray
radiation cannot be ignored during the accretion of the compact object
\citep{rude89,tava93}. Additionally, because the accretion of the IMBH would produce an extremely high X-ray luminosity, the irradiation effect of X-ray flux on the surface of the donor star cannot be neglected for wide-orbit systems.  A fraction of the X-ray flux that the donor star receives is assumed to overcome the gravitational potential energy, and convert
into the kinetic energy of the wind (where the velocity of the wind is thought to be the escape velocity at the donor's surface), which is called the irradiation-driven winds. The stellar wind-loss rate therefore follows the relation
\begin{equation}
L_{\rm X}\frac{\pi R_{\rm d}^{2}}{4\pi a^{2}}f=-\frac{GM_{\rm d}\dot{M}_{\rm wind}}{R_{\rm d}},
\end{equation}
where $R_{\rm d}$ is the radius of the donor star, and $f$ is the wind-driving efficiency.
The irradiation-driven wind loss rate is written as
\begin{equation}
\dot{M}_{\rm ir}=-fL_{\rm X}\frac{R_{\rm d}^{3}}{4 G M_{\rm d} a^{2}}.
\end{equation}
The irradiation-driven winds are assumed to be ejected from the vicinity of the donor star,
and carry away the specific orbital-angular-momentum of the donor star. The mass-loss rate of the donor star
is $\dot{M}_{\rm d}=\dot{M}_{\rm tr}+\dot{M}_{\rm ir}$.

\begin{figure*}
\centering
\begin{tabular}{cc}
\includegraphics[width=0.5\textwidth,trim={10 10 30 30},clip]{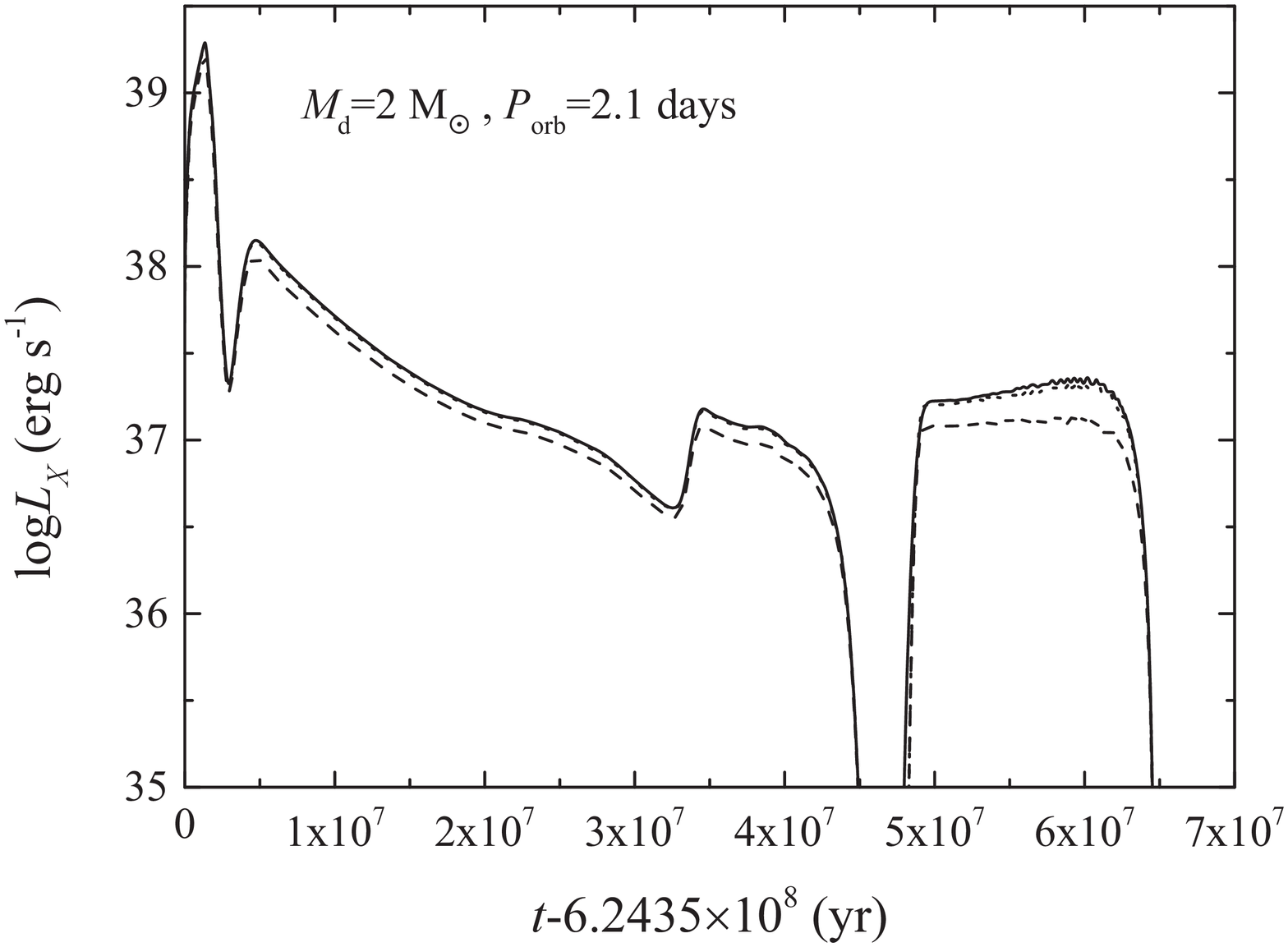} &   \includegraphics[width=0.5\textwidth,trim={10 10 30 30},clip]{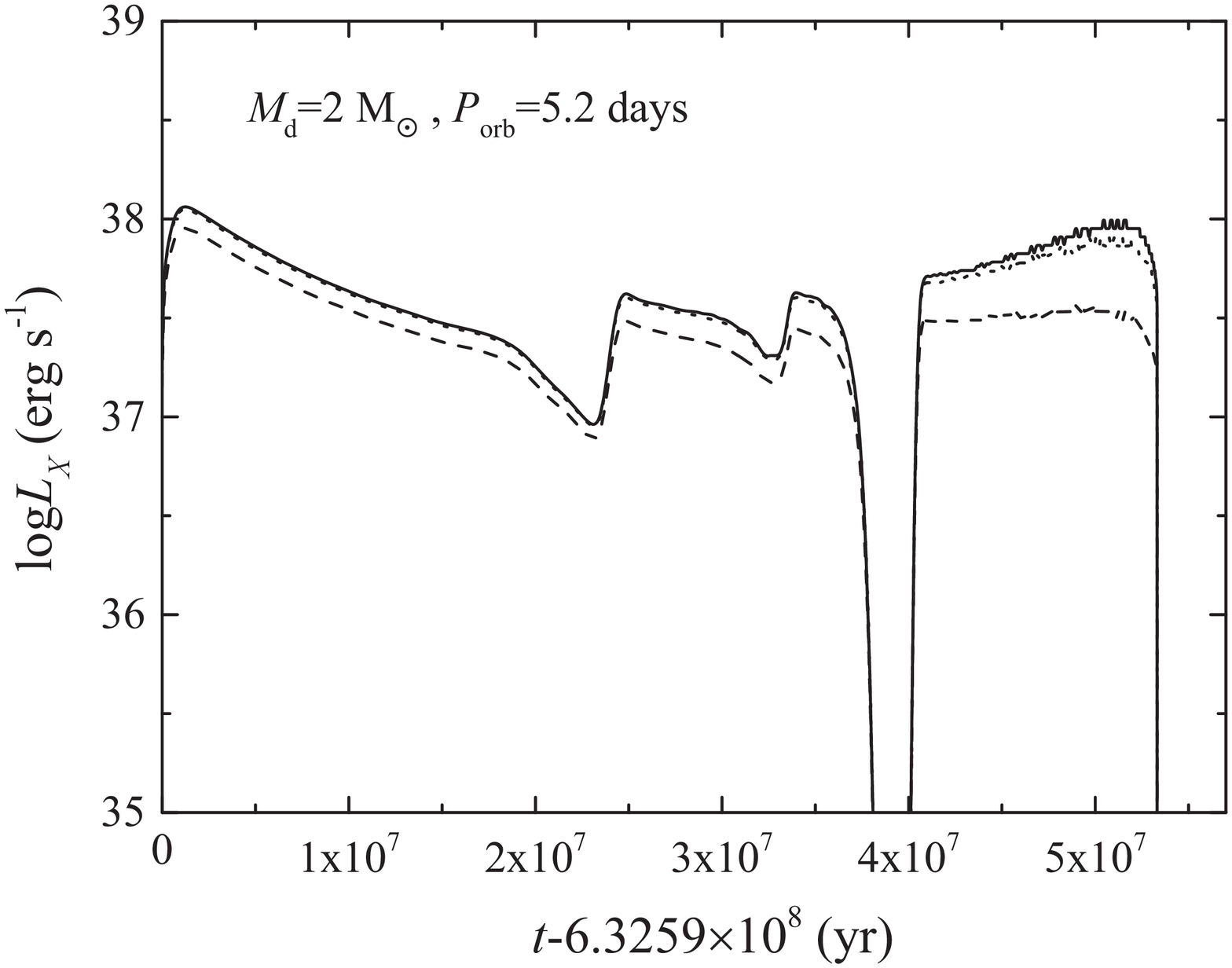} \\
\includegraphics[width=0.5\textwidth,trim={10 10 30 30},clip]{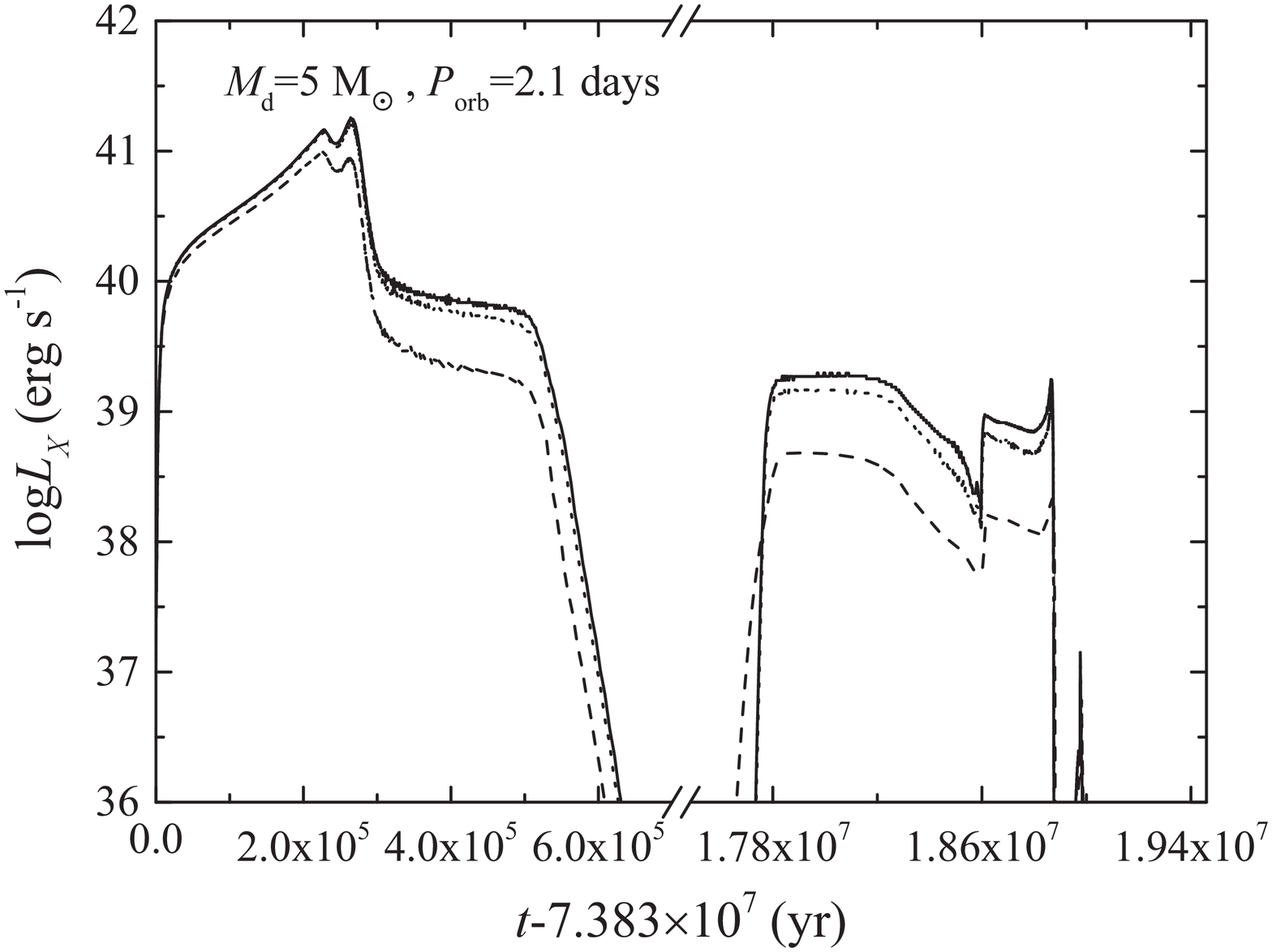} &   \includegraphics[width=0.5\textwidth,trim={10 10 30 30},clip]{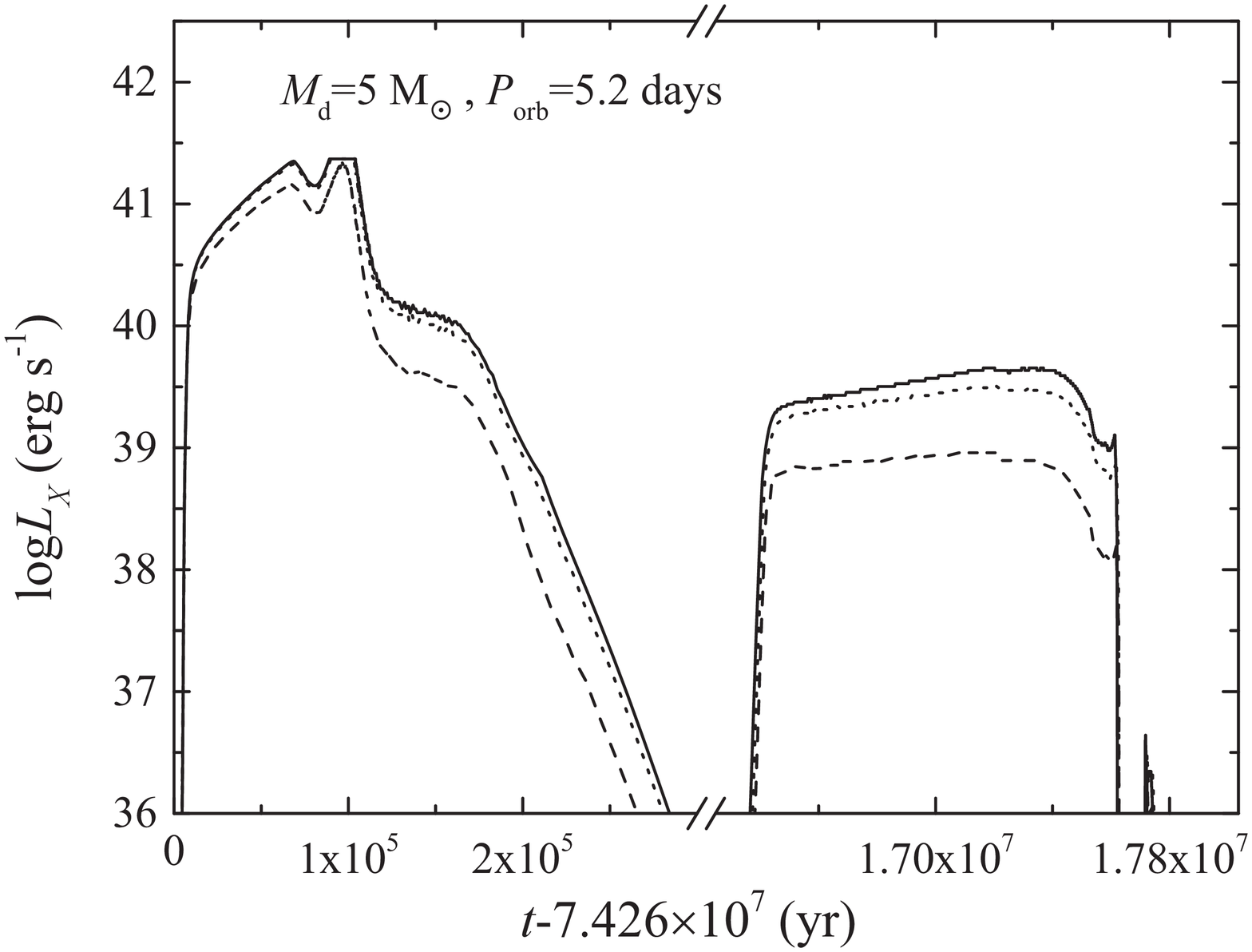} \\
\includegraphics[width=0.5\textwidth,trim={10 10 30 30},clip]{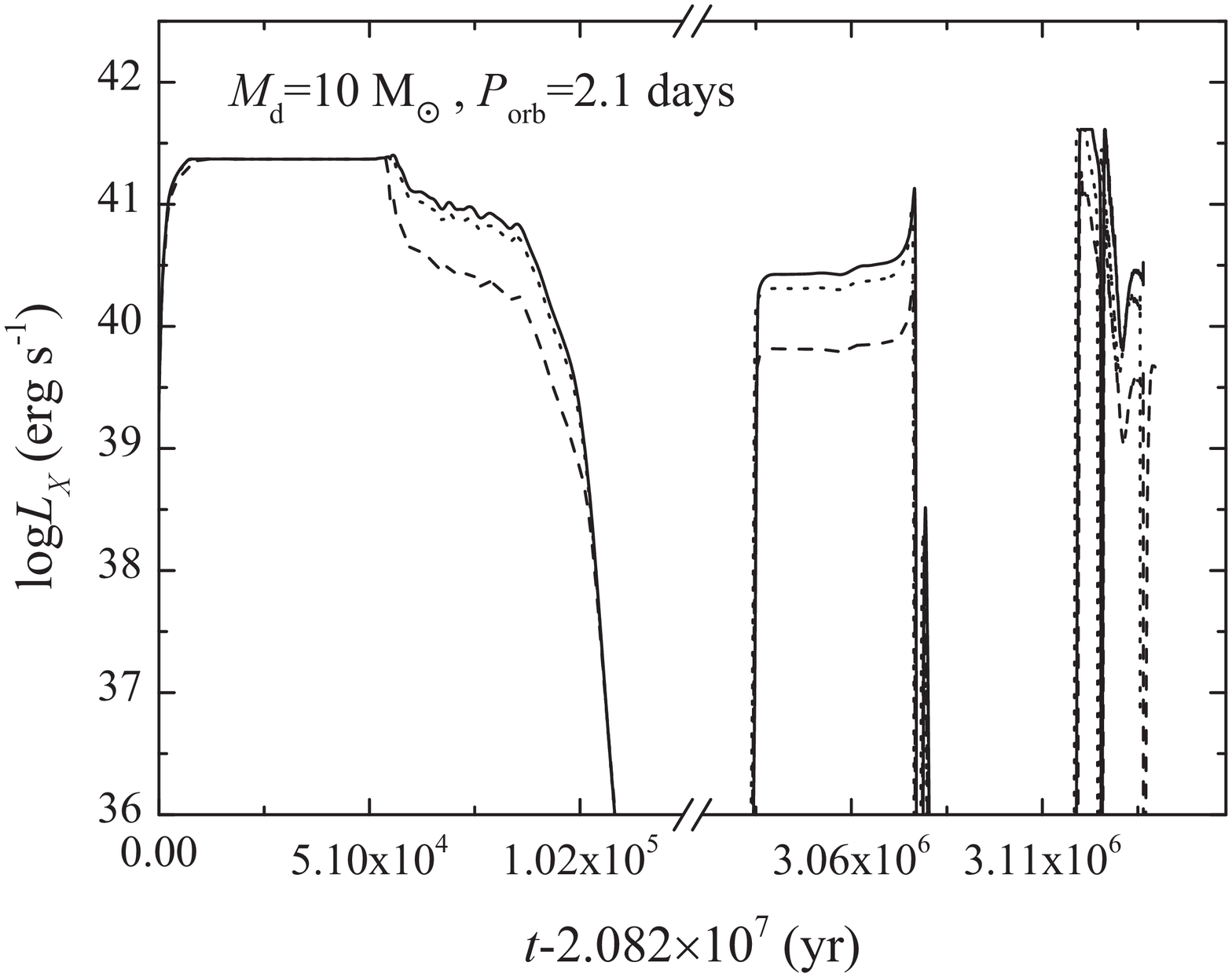} &   \includegraphics[width=0.5\textwidth,trim={10 10 30 30},clip]{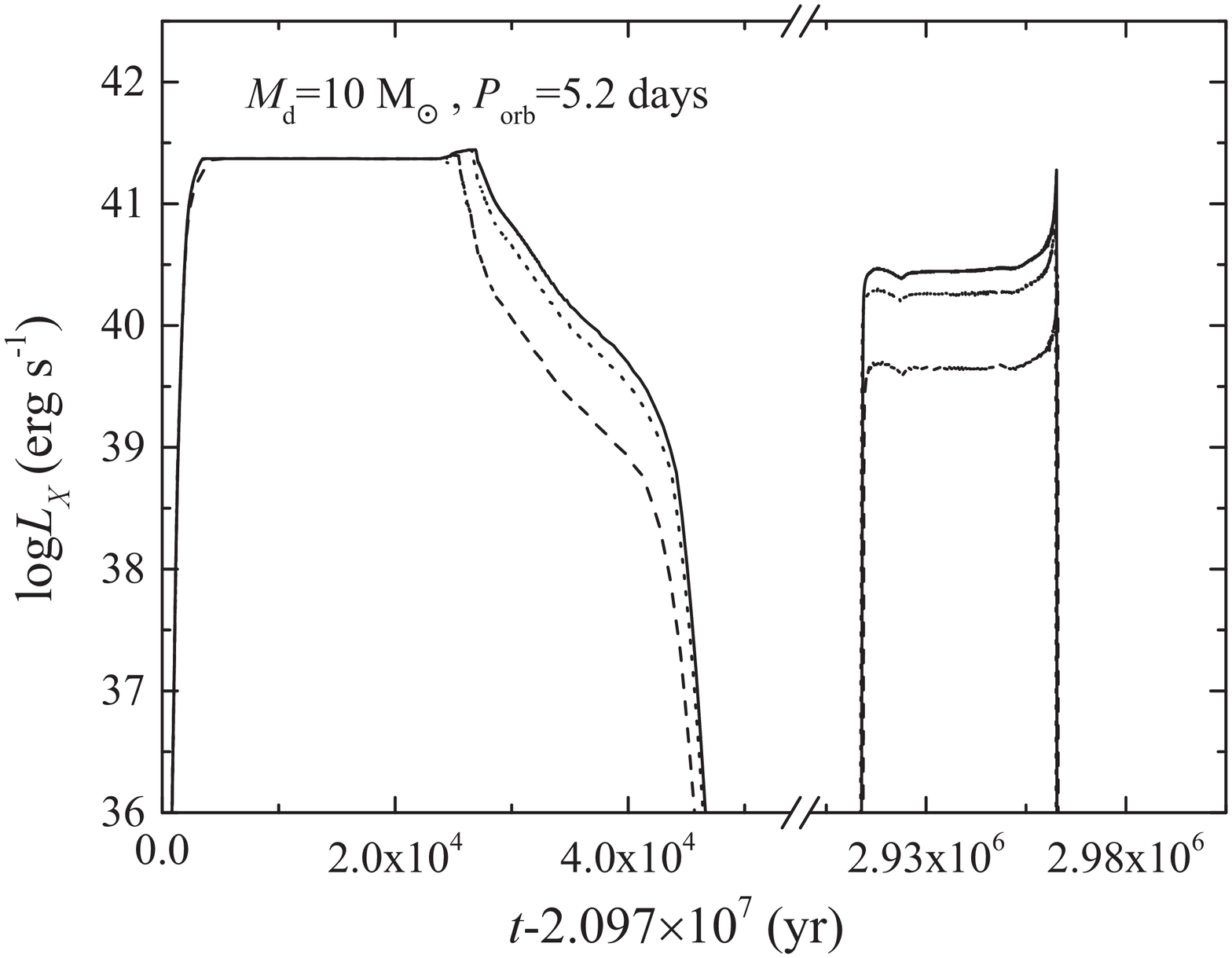} \\
\end{tabular}
\caption{\label{fig:pulses}Same as Figure 1, but for the donor stars with a metallcity $Z=0.001$. The solid, dashed, and dotted curves denote the wind-driving
efficiency $f=0, 0.002$, and 0.0002, respectively. }
\end{figure*}

For donor stars with a metallicity $Z=0.02$, we adopt three wind-driving efficiencies as $f=0.01$, 0.001, and 0. Stars with different metallicities, and different internal structures should have different wind-loss rates, and a modified factor $Z^{1/2}$ is thereby considered \citep{kudr89,hurl00}, i. e. the wind-driven efficiencies for the donor stars with a metallicity $Z=0.001$ are taken to be $f=0.002$, 0.0002, and 0.

\begin{figure*}
\centering
\begin{tabular}{cc}
\includegraphics[width=0.48\textwidth,trim={10 10 30 30},clip]{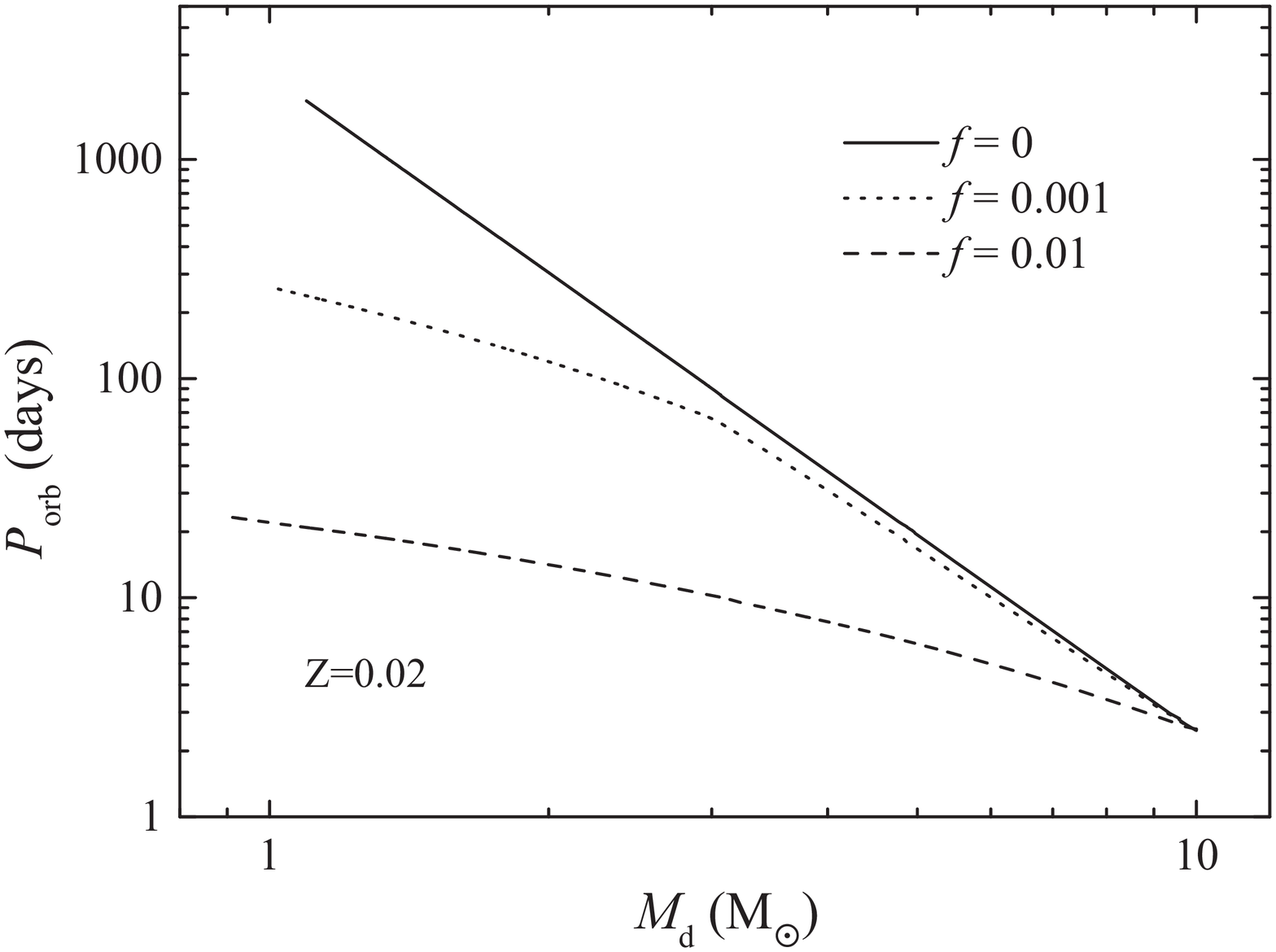} &    \includegraphics[width=0.48\textwidth,trim={10 10 30
30},clip]{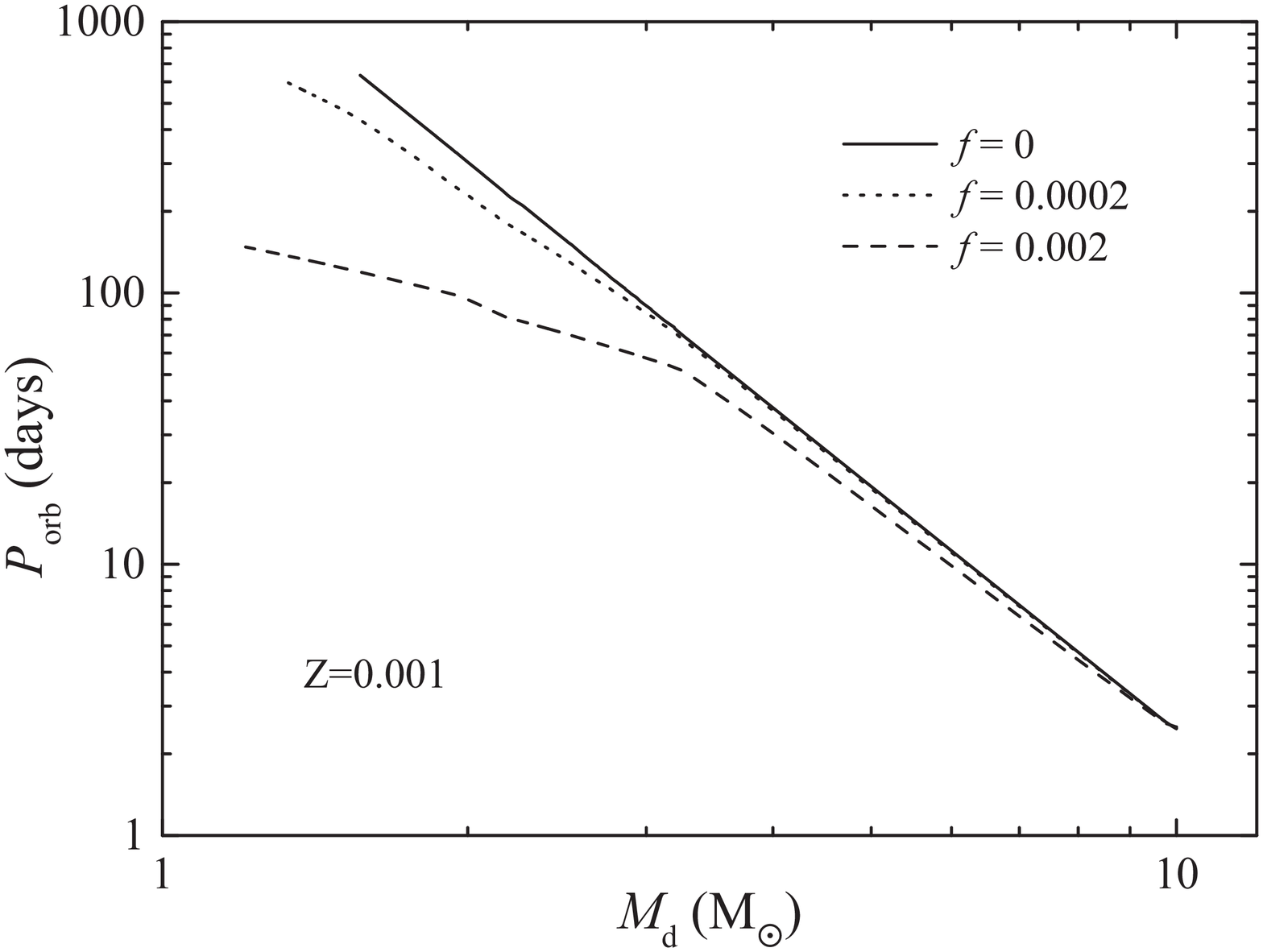} \\
\end{tabular}
\caption{\label{fig:pulses}Evolution of the orbital periods as a function of the donor-star masses
for IMBH X-ray binary with an initial donor-star mass of $10~M_{\odot}$ and an initial orbital period of 2.512 days. The left panel, and the right panel represent a donor-star metallicity of 0.02, and 0.001, respectively. In the left panel, the solid, dashed, and dotted curves denote the wind-driving efficiency $f=0, 0.01$, and 0.001, respectively. In the right panel, the solid, dashed, and dotted curves denote the wind-driving efficiency $f=0, 0.002$, and 0.0002, respectively.}
\end{figure*}

\section{Simulated results}
\subsection{X-ray luminosity}
The evolutionary tracks of IMBH X-ray binaries with different initial donor-star masses and different initial orbital periods in X-ray luminosity versus mass transfer timescale diagrams are plotted in Figures 1 and 2 to illustrate the effect of irradiation-driven winds on the X-ray luminosities of IMBH X-ray binaries. The metallicities of the donor stars are $Z=0.02$, and 0.001 in Figures 1, and 2, respectively. When $Z=0.02$, the effect of a low wind-driving efficiency ($f=0.001$) on the X-ray luminosities of IMBH X-ray binaries with a low-mass donor star is slight, and can be ignored. At $Z=0.001$, the X-ray luminosities of all IMBH X-ray binaries are hardly affected by the irradiation effect for a low wind-driving efficiency ($f=0.0002$). For donor stars with a low-mass of $2~M_{\odot}$, IMBH X-ray binaries are difficult to appear as ULXs (when $Z=0.001$, IMBH X-ray binaries with a short orbital period can appear as ULXs for 1.1 Myr), whereas they are visible as normal X-ray sources at a relatively long timescale of $\sim50-80~\rm Myr$, which depends on the initial orbital periods and metallicities. When the donor stars have an intermediate-mass ($5~M_{\odot}$) or high-mass ($10~M_{\odot}$), a high wind-driving efficiency ($f=0.01$ for $Z=0.02$, and $f=0.002$ for $Z=0.001$) tends produce a low X-ray luminosity. For high-mass donor stars, the effect of irradiation-driven winds can be neglected in the early stage of the mass transfer. All IMBH X-ray binaries have one or two gaps during the mass transfer. The gaps between two mass-transfer stages originate from the thermal-equilibrium deviation of the donor stars after their central hydrogen and helium are exhausted \citep[see also][]{pods03,kalo04,li04,rapp05}.

Table 1 lists main evolutionary parameters of the IMBH X-ray binaries in Figures 1 and 2. To explore the effect of the X-ray irradiation process, two cases that the wind-driving efficiency $f=0$, and 0.01 (0.002 for $Z=0.001$) are considered. When $Z=0.001$, an IMBH X-ray binary with a low donor-star mass and short orbital period can appear as an ULX for $\sim1~\rm Myr$, while the irradiation effect hardly affects the duration of the ULX stage owing to the wind-driving efficiency being a low value of 0.002. All IMBH X-ray binaries with intermediate-mass ($5~M_{\odot}$) or high-mass ($10~M_{\odot}$) donor stars are visible as ULXs at $\bigtriangleup t_{39}\sim0.1-1.8~\rm Myr$. The irradiation effect obviously shortens the duration of the ULX stage for intermediate-mass donor stars, whereas its effect can be ignored for high-mass donor stars. Considering the irradiation effect, the durations of $\bigtriangleup t_{40}$ are also shortened by at least a factor of 2 for donor stars with metallicity of 0.02, while this effect is not evident for donor stars with metallicity of 0.001 owing to the low wind-driving efficiency. Two IMBH binaries with intermediate-mass donor stars ($P_{\rm orb,i}=10~\rm days$ when $Z=0.02$, and $P_{\rm orb,i}=2.1~\rm days$ when $Z=0.001$) do not evolve toward HLXs owing to the irradiation effect. For other systems, the irradiation effect obviously shortens the durations of the HLX stage.

\begin{table*}
\begin{center}
\caption{Some main evolutionary parameters of IMBH binaries in Figures 1 and 2. }
\begin{tabular}{@{}llllllllll@{}}
  \hline\hline\noalign{\smallskip}
$Z$ &$f$&$M_{\rm d,i}$ &$P_{\rm orb,i}$ & $t_{\rm rlof}$   & $P_{\rm rolf}$ & $\bigtriangleup t_{39}$ &  $\bigtriangleup t_{40}$ &
$\bigtriangleup t_{41}$&$L_{\rm X,max}$\\
& &($\rm M_{\odot}$)  &(days)& (Myr)&(days) &(Myr)&(Myr)&($10^{4}$ yr)&($10^{41}\rm erg\,s^{-1}$)\\
 \hline\noalign{\smallskip}
         &0    &2.0& 3.0 & 986.89  & 2.72  &  0 &  0 & 0  & 0.001      \\
         &0.01 &2.0& 3.0 & 986.89  & 2.72  &  0 &  0 & 0  & 0.0006      \\
         & 0  &2.0& 10.0 & 1016.09  & 9.96  &  0 &  0 & 0  & 0.002      \\
         &0.01&2.0& 10.0 & 1016.09  & 9.96  &  0 & 0  &  0 &  0.0003    \\
         &0    &5.0& 3.0 & 84.53   & 2.94  & 1.756  & 0.683  &  0 &  0.5     \\
0.02     &0.01 &5.0& 3.0 & 84.53   & 2.94  & 0.744  & 0.160  &  0 &  0.2     \\
         & 0  &5.0& 10.0 & 85.41   & 9.99  & 1.539  & 0.426  & 3.87  &2.4       \\
         &0.01&5.0& 10.0 & 85.41   & 9.99  & 0.368  & 0.106  & 0     &0.4     \\
         &0   &10.0& 3.0 & 19.70   & 2.97  & 0.155  & 0.149  & 6.60  &2.7       \\
         &0.01&10.0& 3.0 & 19.70   & 2.97  & 0.140  & 0.061  & 5.19  &2.5       \\
         & 0  &10.0&10.0 & 19.85  &  9.997 & 0.122  & 0.086  & 3.32  &2.8       \\
         &0.01&10.0&10.0 & 19.85  &  9.997 & 0.053  & 0.025  & 1.99  &2.4      \\
\hline\noalign{\smallskip}
         &0     &2.0& 2.1 & 624.35  &1.77   & 1.104  & 0  &  0 & 0.02      \\
         &0.002 &2.0& 2.1 & 624.35  &1.77   & 1.104  & 0  &  0 & 0.02      \\
         & 0    &2.0& 5.2 & 632.59  &5.13   & 0  &  0 & 0  &  0.001     \\
         &0.002 &2.0& 5.2 & 632.59  &5.13   & 0  &  0 & 0  &  0.001    \\
         &0     &5.0& 2.1 & 73.83   & 2.01  & 1.127 &0.304   & 7.60  &  1.8     \\
0.001    &0.002 &5.0& 2.1 & 73.83   & 2.01  & 0.514 &0.266   & 0     &  1.0     \\
         & 0    &5.0& 5.2 & 74.26   & 5.18  & 1.427 &0.156   & 6.91  &  2.3     \\
         &0.002 &5.0& 5.2 & 74.26   & 5.18  & 0.176 &0.106   & 4.32  &  2.2    \\
         &0     &10.0& 2.1 & 20.82  & 2.05  & 0.150 &0.143   & 7.52  &  4.1     \\
         &0.002 &10.0& 2.1 & 20.82  & 2.05  & 0.147 &0.100   & 5.71  &  2.4     \\
         & 0    &10.0& 5.2 & 20.97  & 5.19  & 0.088 &0.081   & 2.67  &  2.8     \\
         &0.002 &10.0& 5.2 & 20.97  & 5.19  & 0.084 &0.030   & 2.37  &  2.7    \\
\noalign{\smallskip}\hline
\end{tabular}
\tablenotetext{}{}\\ {Note. The columns list (in order): metallicity, wind-driving efficiency, the initial-donor mass, initial orbital period, stellar age at RLOF, orbital period at ROLF, durations that the X-ray luminosity exceeds $10^{39}~\rm erg\,s^{-1}$, $10^{40}~\rm erg\,s^{-1}$, $10^{41}~\rm erg\,s^{-1}$, maximum X-ray luminosity.  }
\end{center}
\end{table*}

\subsection{Orbital evolution and GW sources}
During the evolution of IMBH X-ray binaries, there is a bifurcation period, which is defined as the longest initial orbital period of a binary evolving into ultra-compact X-ray binary within the Hubble timescale \citep{sluy05a,sluy05b}.
The material transfer from the less massive donor star to the more massive IMBH would cause orbital periods to gradually increase,
while the loss of angular-momentum would produce an inverse tendency. The competition between orbital expansion
and orbital shrinkage results in a bifurcation period \citep{liu17,liu21}, which plays an important role in determining the final fate of a binary system including an accreting object \citep{pods02,ma09,chen16,jia16}.

Figure 3 shows evolutionary examples of IMBH binaries with initial orbital periods longer than the bifurcation period
in the $P_{\rm orb}-M_{\rm d}$ diagrams.  It is clear that the irradiation-driven winds strongly affect the orbital evolution of IMBH X-ray binaries. In particular, the
irradiation-driven winds are ejected from the vicinity of the donor star, and extract the specific orbital-angular-momentum expressed as follows
\begin{equation}
j_{\rm ir}=\frac{JM_{\rm bh}}{(M_{\rm bh}+M_{\rm d})M_{\rm d}},
\end{equation}
where $J$ is the total orbital-angular-momentum of the binary. The isotropic winds during the super-Eddington accretion
should be ejected from the vicinity of the IMBH, and the specific orbital-angular-momentum is thus given by
\begin{equation}
j_{\rm is}=\frac{JM_{\rm d}}{(M_{\rm bh}+M_{\rm d})M_{\rm bh}}.
\end{equation}
The ratio of these two values $j_{\rm ir}/j_{\rm is}=(M_{\rm bh}/M_{\rm d})^{2}\approx(1000/10)^{2}=10^{4}$ , i.e. the efficiency of the extraction of angular momentum by the irradiation-driven winds is four orders of magnitude greater than that of the extraction of angular momentum by the isotropic winds. The irradiation-driven winds are therefore an important mechanism of angular-momentum-loss that affects the orbital evolution of IMBH X-ray binaries. When the donor-stars (with a metallicity $Z=0.02$) evolve into a mass $M_{\rm d}\approx 1~M_{\odot}$, the orbital periods are approximately 20, 200, and 2000 days for a wind-driving efficiency
$f=0.01,0.001$, and 0, respectively. For donor stars with low metallicity ($Z=0.001$), the increasing rates of the orbital periods
are obviously higher than those of the donor stars having $Z=0.02$ when the irradiation effect is included. This difference likely originates from the relatively low wind-driving efficiency that we adopted for the donor stars with low metallicity.

\begin{figure*}
\centering
\begin{tabular}{cc}
\includegraphics[width=0.48\textwidth,trim={10 10 30 30},clip]{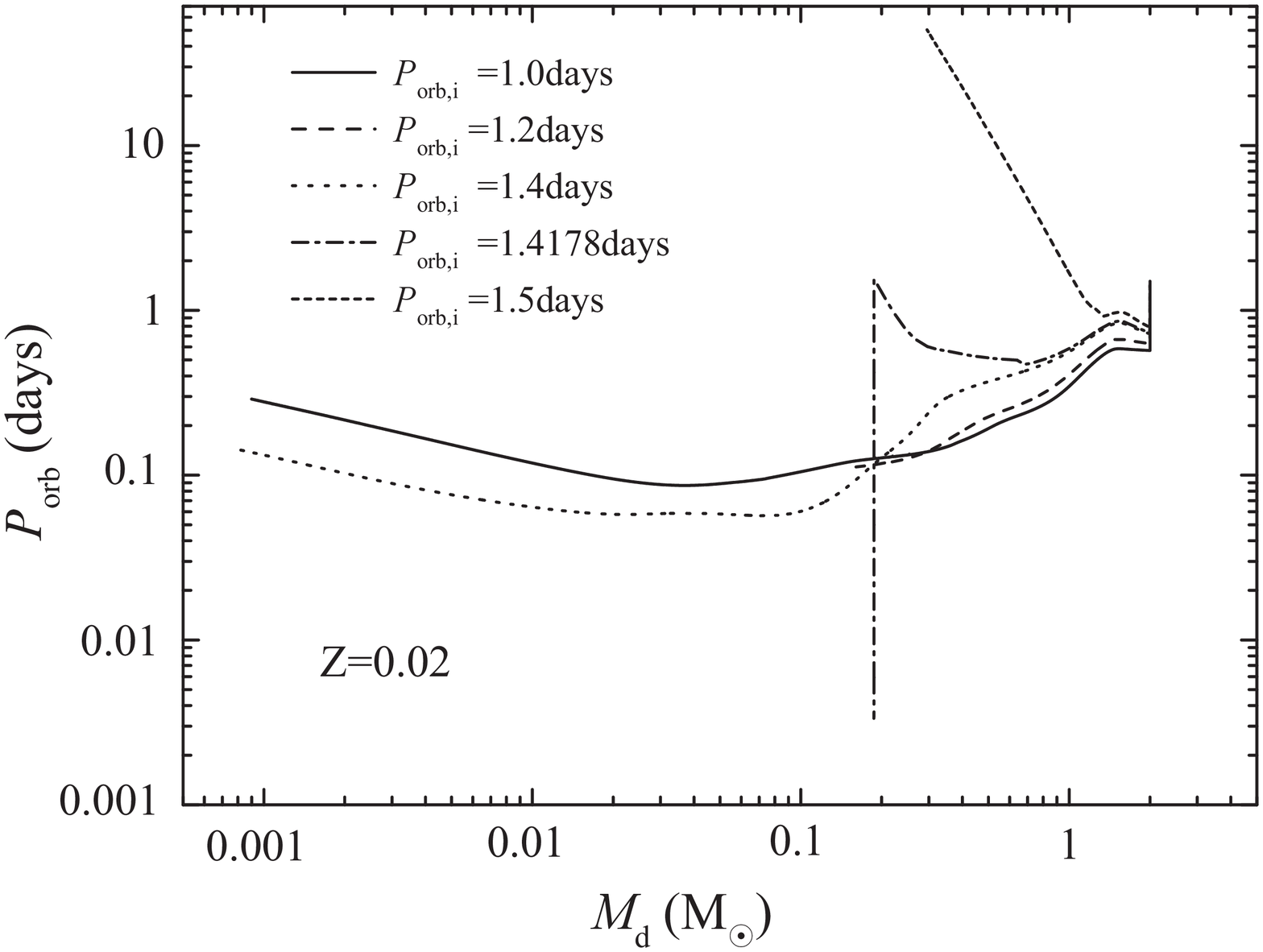} &    \includegraphics[width=0.48\textwidth,trim={10 10 30
30},clip]{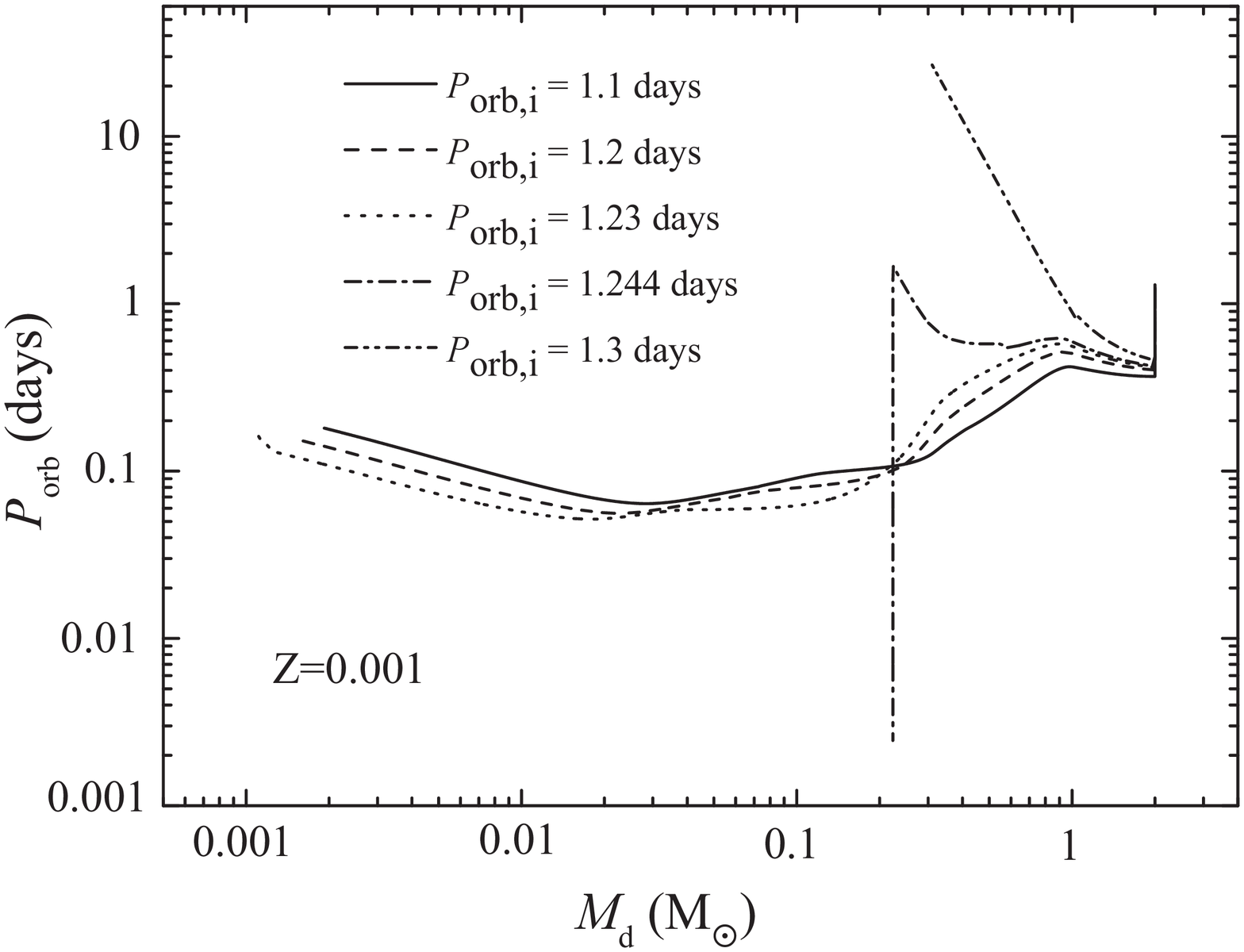} \\
\end{tabular}
\caption{\label{fig:pulses}Evolution of the orbital periods as a function of the donor-star masses
for IMBH X-ray binaries consisting of an IMBH with an initial mass of $1000~M_{\odot}$ and a donor star with an initial mass
of $2~M_{\odot}$ for different initial orbital periods. The left panel, and the right panel represent a donor-star metallicity of
0.02, and 0.001, respectively. The wind-driving
efficiency $f=0.01$, and 0.002 in the left, and right panels, respectively. }
\end{figure*}

If the initial orbital periods of IMBH binaries are shorter than the bifurcation period, these sources would evolve toward compact binary systems, which are potential Laser Interferometer Space Antenna (LISA) sources \citep{port04b,chen20}. Figure 4 presents the evolution of IMBH binaries with a $2~M_{\odot}$ donor star and short orbital periods on a $P_{\rm orb}-M_{\rm d}$ diagram. The bifurcation periods are 1.4178, and 1.244 days for a $2~M_{\odot}$ donor star with metallicity of 0.02, and 0.001, respectively. The IMBH binaries with initial orbital periods much shorter than the bifurcation period tend to become compact binaries with relatively long orbital periods \citep[see also][]{pods02,chen16,chen20}. When the initial orbital period equals the bifurcation period, the donor stars decouple their Roche lobes at $P_{\rm orb}\approx1.5~\rm days$, and the IMBH X-ray binaries evolve into detached binaries with an IMBH and a He core ($\sim0.2~M_{\odot}$). Subsequently, the He core first evolves into a He white dwarf (WD) after a contraction phase, and then enters a cooling stage \citep{istr14}. Owing to the strong gravitational radiation, the orbits of the IMBH binaries rapidly shrink, and the IMBH binaries eventually evolve into ultracompact IMBH-WD binaries with orbital periods of 4.3 minutes in the Hubble timescale. For detached binaries consisting of a neutron star (NS) and a WD, only the systems with an initial orbital period less than $7-9$ hours can become visible LISA sources \citep{taur18,chen20b}. This difference arises from the masses of IMBHs being approximately three orders of magnitude higher than those of NSs, which results in a rate of angular-momentum-loss via gravitational radiation that is at least seven orders of magnitude higher than that in NS systems \citep[see also][]{chen20}. Our simulations also show that the wind-driving efficiency hardly affect the evolutionary tracks in Figure 4 (see also subsection 3.3).

\begin{figure*}
\centering
\begin{tabular}{cc}
\includegraphics[width=0.48\textwidth,trim={10 10 30 30},clip]{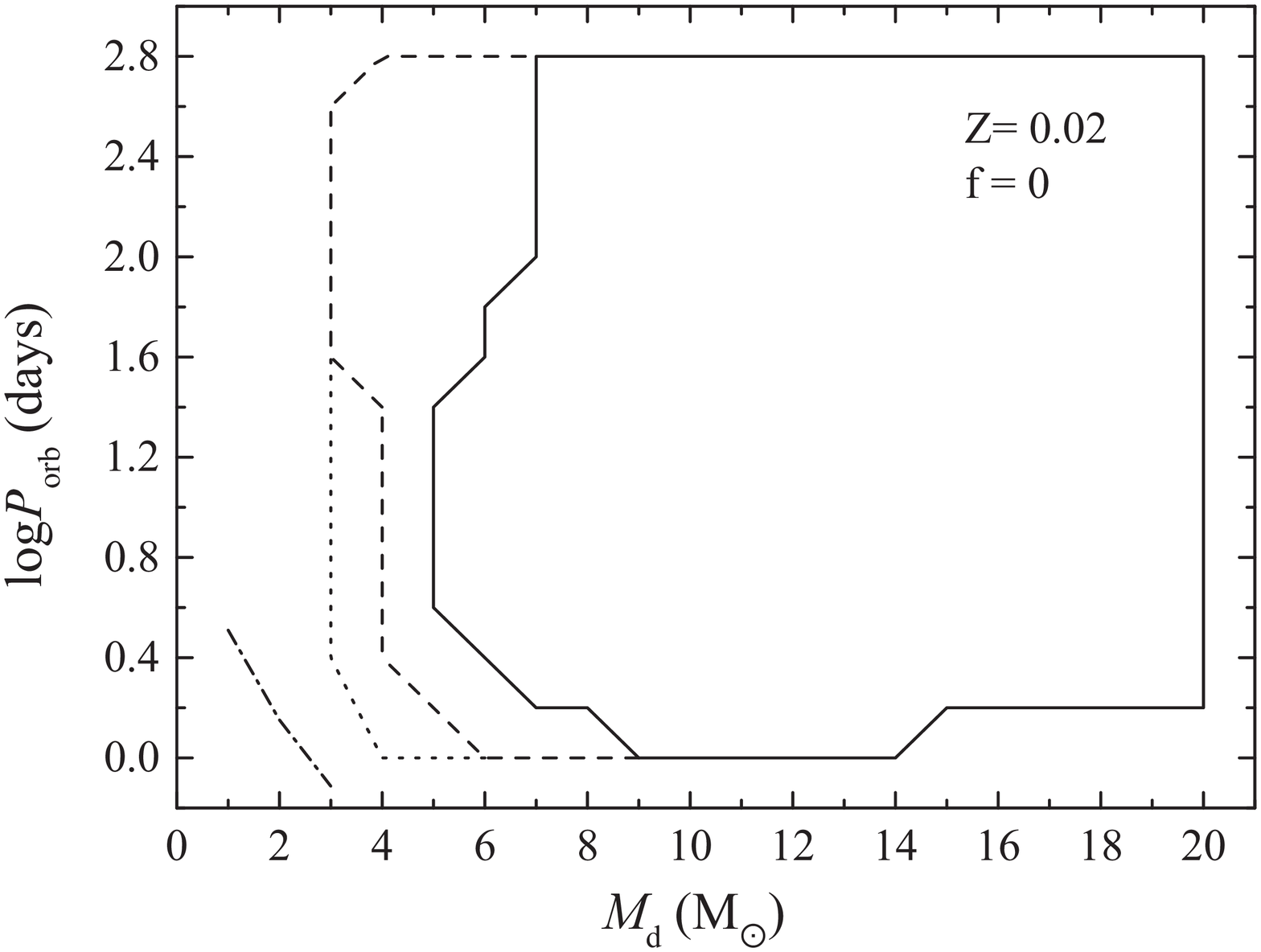} &    \includegraphics[width=0.48\textwidth,trim={10 10 30   30},clip]{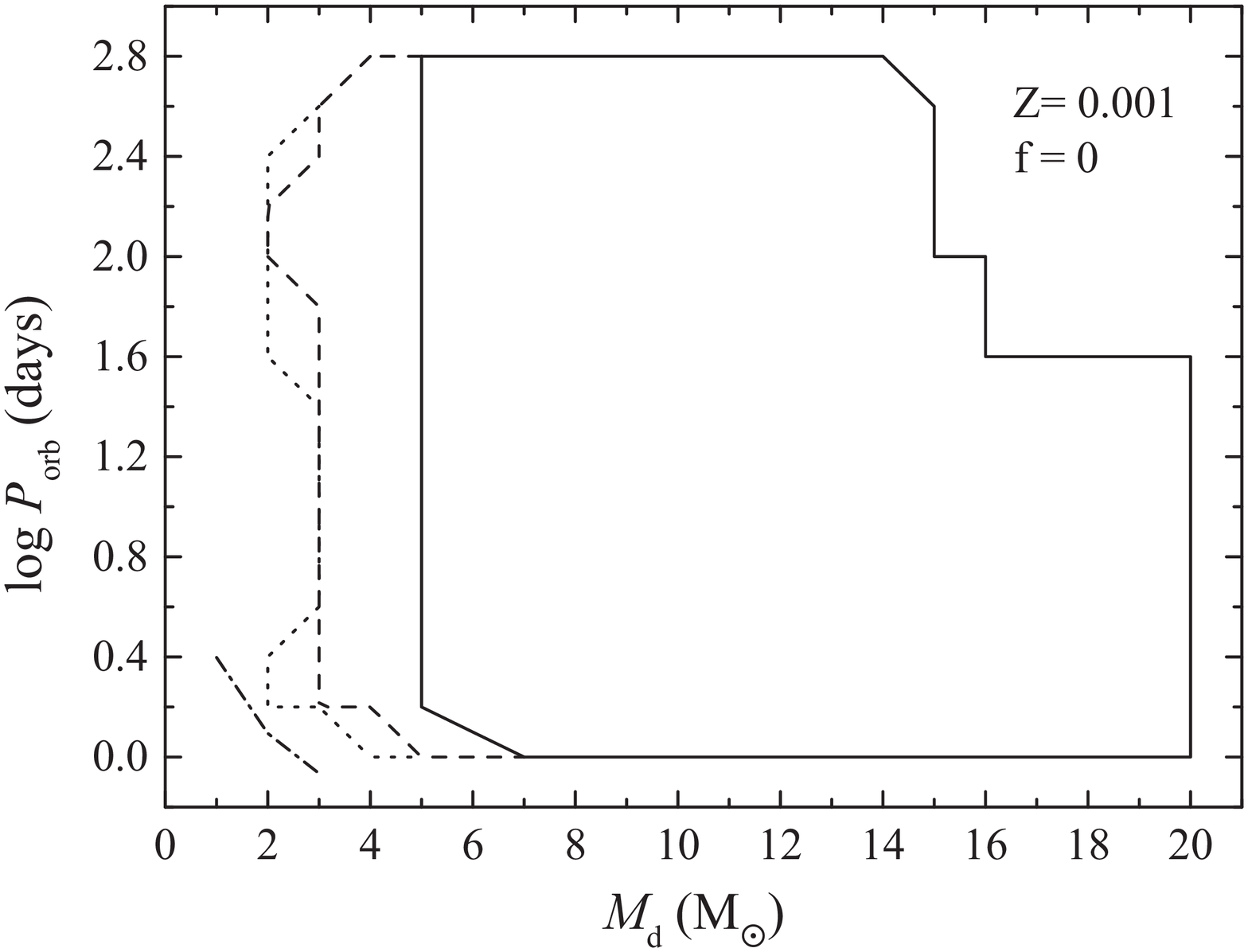} \\
\includegraphics[width=0.48\textwidth,trim={10 10 30 30},clip]{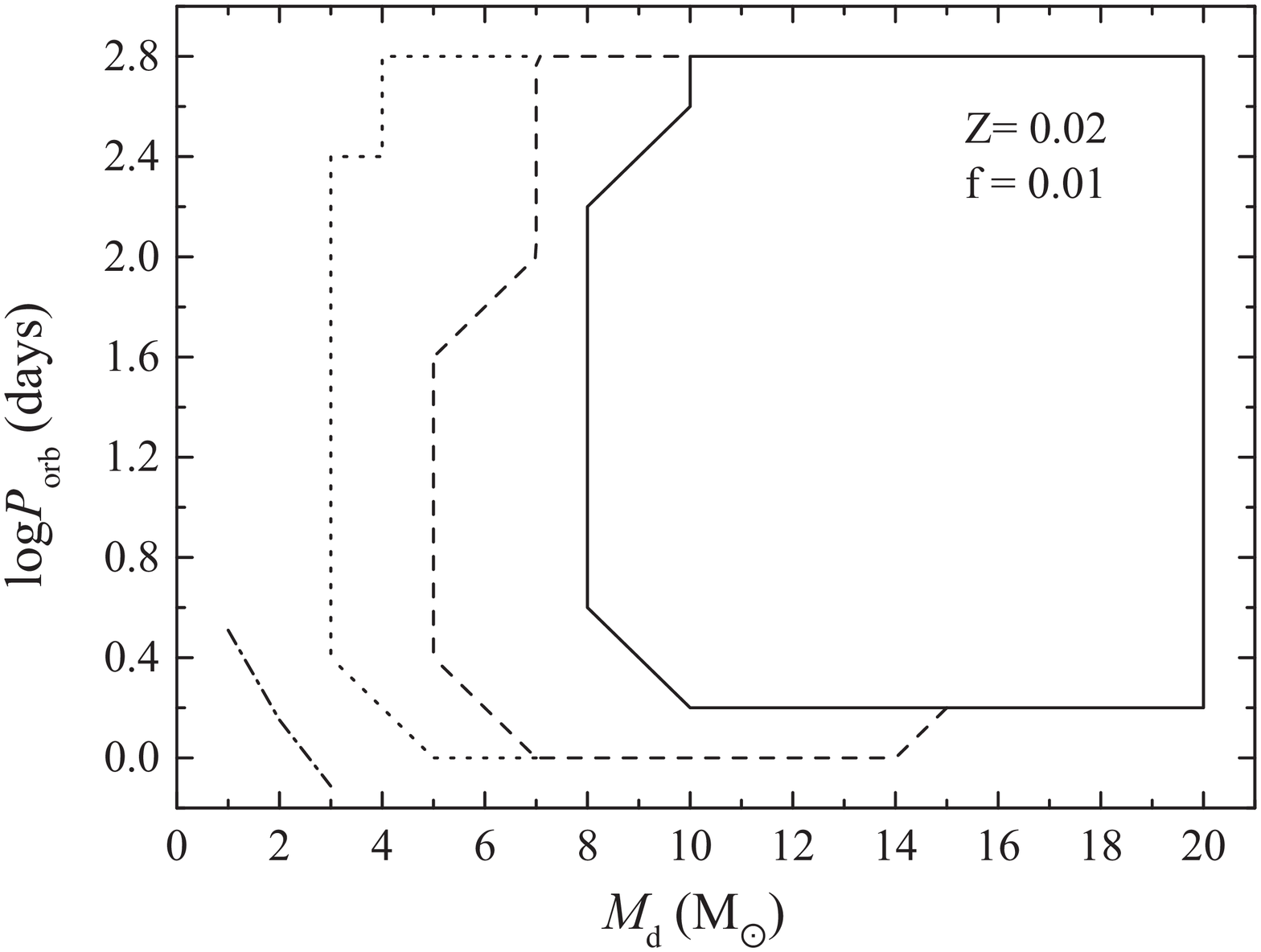} &    \includegraphics[width=0.48\textwidth,trim={10 10 30   30},clip]{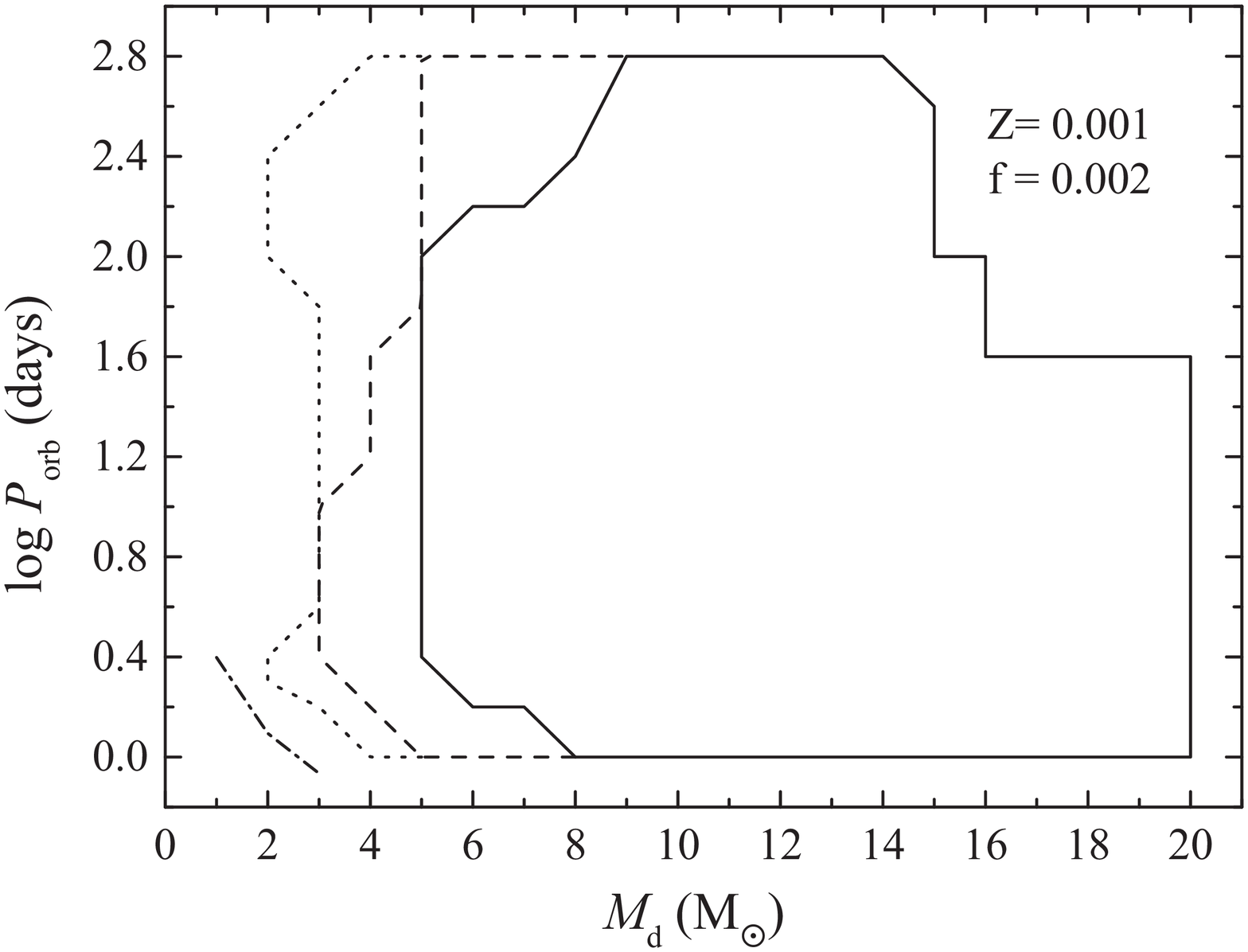} \\
\end{tabular}
\caption{\label{fig:pulses}
Parameter space of the IMBH binaries with different maximum X-ray luminosities in the initial orbital period vs. initial
donor-star mass diagram. The left, and the right panels represent a donor-star metallicity $Z=0.02$, and 0.001, respectively. The upper panels correspond to the wind-driving efficiency $f=0$, while the wind-driving efficiency $f=0.01$, and 0.002 in the left and right panels on the bottom, respectively. The dashed-dotted curves denote the bifurcation period. The contours enclosed by the solid, dashed, and dotted curves describe the maximum X-ray luminosities $L_{\rm X}>10^{41}~\rm erg\,s^{-1}$, $L_{\rm X}>10^{40}~\rm erg\,s^{-1}$, and $L_{\rm X}>10^{39}~\rm erg\,s^{-1}$, respectively.  }
\end{figure*}

\subsection{Initial parameter of ULXs}
The orbital period versus donor-star mass plane (${\rm log}(P_{\rm orb, i}/\rm days)=0-3$, and $M_{\rm d,i}=1-20~M_{\odot}$, where $P_{\rm orb, i}$, and $M_{\rm d,i}$ are respectively the initial orbital period and the initial donor-star mass) is uniformly divided into $20\times15$ discrete grids, in which we diagnose whether the IMBH binaries evolve toward ULXs when the wind-driving efficiency $f=0$, and 0.01 (or 0.002). Figure 5 shows the initial contours for ULXs with different maximum X-ray luminosities in the ${\rm log}P_{\rm orb, i}-M_{\rm d,i}$ plane. It is clear that IMBH binaries with a low donor-star mass ($1-2~M_{\odot}$) and a short orbital period ($\sim1-2$ days) cannot evolve toward ULXs. In Figure 5, the dashed-dotted curves denote the bifurcation periods. All IMBH binaries with an initial orbital period shorter than the bifurcation periods only appear as normal X-ray sources. Our simulations also show that the bifurcation periods are hardly affected by the wind-driving efficiency. The X-ray luminosities of these systems are relatively low (only $\sim 10^{38}~\rm erg\,s^{-1}$), which results in a small irradiation-driven wind loss rate and thus a small angular-momentum-loss rate. Both left and right panels have four grids under the bifurcation periods. This implies that about 1\% of IMBH binaries in our simulated parameter space can evolve toward compact X-ray sources, with some appearing as low-frequency GW sources detectable by LISA \citep{chen20}.

When considering the irradiation-driven winds, the left boundaries of the parameter spaces with $L_{\rm X}>10^{40}~\rm erg\,s^{-1}$ and $L_{\rm X}>10^{41}~\rm erg\,s^{-1}$ for $Z=0.02$ move toward the high donor-star masses, whereas the bottom and upper boundaries move toward the long orbital periods and short orbital periods, respectively. When $Z=0.001$, the shift directions of the upper (long orbital periods) and bottom (short orbital periods) parts of left boundaries with $L_{\rm X}>10^{40}~\rm erg\,s^{-1}$ and $L_{\rm X}>10^{41}~\rm erg\,s^{-1}$ are similar to those for $Z=0.02$. It is clear that the irradiation-driven winds cannot readily affect the initial parameter of ULXs, whereas it shrinks the initial parameter space of IMBH binaries that evolve toward ULXs with high luminosity ($L_{\rm X}>10^{40}~\rm erg\,s^{-1}$) and HLXs. Considering the irradiation effect, there are 220 and 223 grids within the contours with $L_{\rm X}>10^{40}~\rm erg\,s^{-1}$ for $Z=0.02$, and 0.001, respectively, while these two numbers increase to be 252, and 240 without the irradiation effect. Assuming that initial donor-star masses and initial orbital periods satisfy a uniform distribution, 58\% (174 grids) and 65\% (195 grids) of IMBH binaries can evolve toward HLXs when $Z=0.02$, and 0.001, respectively. If the irradiation effect is not included, these two percentages respectively increase to be 72\% (215 grids) and 70\% (209 grids).

\section{Discussion}
IMBH X-ray binaries are broadly important in astrophysics in that (1)
they are important candidates of ULXs and HLXs \citep{hopm04,port04b,li04,patr06}; and (2) some IMBH X-ray binaries with compact orbits are potential GW sources that would be visible to LISA, TianQin, and Taiji \citep{port04b,chen20}.

Considering the irradiation effect, our simulations indicate that the initial parameter space of IMBH X-ray binaries evolving toward HLXs tend to shrink, especially for the donor stars with a metallicity of $Z=0.02$. Meanwhile, the durations of the HLX stage shorten. In an X-ray and optical study, \cite{sutt12} found that the X-ray spectrum properties of eight ULX candidates are consistent with the sub-Eddington hard state, providing evidences of the existence of IMBHs with masses in the range of $10^{3}-10^{4}~M_{\odot}$. However, only two objects can be classed as HLX candidates in at least one observation \citep{sutt12}, implying a very small number of HLXs identified so far. The irradiation effect may be responsible for the phenomenon that most IMBH X-ray binaries do not readily appear as HLXs at any stage of their life .

It strongly depends on the X-ray luminosity whether the irradiation effect plays a key role in determining the evolution of X-ray binaries. Neglecting the irradiation effect, \cite{li04} found that IMBH and stellar-mass BH X-ray binaries are similar in terms of their donor stars and binary orbits. After the Eddington accretion rate is included, the X-ray luminosities of IMBH X-ray binaries are approximately two orders of magnitude higher than those of stellar-mass BH X-ray binaries, yielding high loss rates of irradiation-driven winds and high rates of angular-momentum-loss. Therefore, the influence of the irradiation effect on IMBH X-ray binaries would obviously exceed that on stellar-mass BH X-ray binaries.

If the initial orbital periods are less than the bifurcation period, the IMBH X-ray binaries would evolve toward compact X-ray sources, and be visible as continuous low-frequency GW sources \citep{chen20}. For the fine-tuning of the initial orbital periods that are near the bifurcation period, the final products of IMBH X-ray binaries are detached IMBH binaries including a low-mass He WD ($\sim0.2~M_{\odot}$, see also Figure 4). With the spiraling-in of the WD, the IMBH-WD binary will emit low-frequency GWs, and appears as a LISA source. Once the WD penetrates the tidal radius of the IMBH, it will yield accretion-driven flares from the tidal disruption of the WD by the IMBH, which may account for gamma-ray burst (GRB) GRB060218 \citep{shch13}, and two fast X-ray transients XT1 and XT2 in the Chandra Deep Field data \citep{peng19}. However, the influence of the irradiation effect on these processes can be ignored because their progenitors are normal X-ray sources.

If the initial donor-star masses are greater than $10~M_{\odot}$, IMBH X-ray binaries would evolve toward IMBH-NS binaries, which may be discovered by deeper surveys or next-generation radio telescopes such as the Square Kilometer Array \citep{patr05}. About $9.5\%$ of IMBH binaries are disrupted in their supernova explosions, which results in an isolated IMBH and an isolated radio pulsar \citep{port10}. Of the surviving IMBH-NS systems, only 0.6\% merge to appears as GW sources because most systems should have relatively long orbital periods (of at least 10 days) \citep{hopm05}. Because the irradiation effect would efficiently drive an angular-momentum loss for ULXs, producing relatively compact orbits, the fraction of disrupted IMBH binaries and IMBH-NS GW sources should decrease and increase, respectively.

Observational constraints proposed a space density for globular clusters as $n_{\rm GC}\approx 4~\rm Mpc^{-3}$ \citep{brod06,rami09}. \cite{hopm04} found that IMBHs capture companions and then successfully circularize at a rate
$\Gamma\approx 5\times 10^{-8}~\rm yr^{-1}$ in globular clusters. Using the initial mass function, we can roughly estimate the probability that an IMBH captures a donor stars with a mass in the range from $i$ to $j(M_{\odot})$ as
\begin{equation}
P_{{\rm c},i-j}=\int_{i}^{j}a\xi(M){\rm d}M,
\end{equation}
where $\xi(M){\rm d}M$ is the probability that the mass of a star is in the range from $M$ to $M+{\rm d}M$ \citep{krou93}.
Assuming that the masses of stars in globular clusters are in the range of $1-20~M_{\odot}$,  i. e. $P_{c,1-20}=1$,
we have $P_{c,1-3}=0.85$, and $P_{c,4-20}=0.15$. Assuming that the initial orbital periods of IMBH binaries obey a uniform distribution, the probability that the IMBH binaries with a donor-star mass in the range from $i$ to $j(M_{\odot})$ appear as ULXs can be approximately estimated as $P_{\rm ULX,1-3}=0.24$, and $P_{\rm ULX,4-20}=0.98$ for $Z=0.02$ and when $f=0.01$; and $P_{\rm ULX,1-3}=0.44$, and $P_{\rm ULX,4-20}=0.89$ for $Z=0.001$ and $f=0.002$ (where the probabilities are calculated by counting the grid numbers within the contours). Therefore, the estimated observation rate of ULXs in the globular clusters (taking $Z=0.02$) is roughly
\begin{eqnarray}
n_{\rm GC,ULX}=& n_{\rm GC}\Gamma(P_{c,1-3}P_{\rm ULX,1-3}+P_{c,4-20}P_{\rm ULX,4-20})\nonumber\\
&\approx 70~\rm Gpc^{-3}yr^{-1}.
\end{eqnarray}
According to the results of section 3.3, the irradiation effect can not influence the observation rate of ULXs in globular clusters. However, the estimated observation rate of HLXs would obviously decline if the irradiation effect is included.

\section{Summary}
As the transitional population between the stellar-mass BH and supermassive BH, IMBHs have
generally been thought to reside in globular clusters, young dense clusters, and dwarf galaxies \citep{gurk04,port04a,baum17}.
In young dense clusters, main sequence stars could spiral into IMBHs via exchange encounters or
tidal capture, and then feed material to the IMBHs via the Roche-lobe overflow after the orbits are circularized \citep{hopm04}.
Considering irradiation-driven winds produced at the surface of the donor star by strong X-ray flux from the accretion disk around the IMBH, we modeled the evolution of a large number of IMBH X-ray binaries. Our main conclusions are as follows:

1. The effect of the irradiation process with a low wind-driving efficiency ($f=0.001$ for $Z=0.02$, and $f=0.0002$ for $Z=0.001$) on the X-ray luminosities of IMBH X-ray binaries can be ignored.

2. A relatively high wind-driving efficiency ($f=0.01$ for $Z=0.02$, and $f=0.002$ for $Z=0.001$) obviously shortens the duration of the ULX stage of IMBH X-ray binaries with an intermediate-mass donor star. However, this effect is very slight for high-mass donor stars. When $Z=0.02$, the irradiation effect shortens the durations of $\bigtriangleup t_{40}$ by a factor at least 2, while this effect is not obvious for the donor stars with $Z=0.001$.

3. Irradiation-driven winds play an important role in determining the orbital evolution of IMBH X-ray binaries. Because the irradiation-driven winds are ejected from the vicinity of the donor star, the rate of angular-momentum-loss is four orders of magnitude higher than that of isotropic winds induced by the super-Eddington accretion for the same mass-loss rate. Therefore, the final orbital periods of IMBH X-ray binaries are obviously shorter than those without the irradiation effect.

4. The irradiation effect can not affect the initial parameter space of IMBH binaries that evolve toward ULXs, while it obviously shrinks those evolving toward ULXs with high luminosity and HLXs. Taking $f=0.01$ ($Z=0.02$) and 0.002 ($Z=0.001$), 58\% and 65\% of IMBH binaries with a metallicity $Z=0.02$, and 0.001 on the initial orbital periods versus initial donor-star masses plane have an opportunity to evolve toward HLXs, respectively. If the irradiation effect is not included, these two fractions increase to be 72\% and 70\%.

5. A low donor-star mass ($1-2~M_{\odot}$) or short orbital period ($\sim 1-2$ days) tends to give rise to the emergence of normal X-ray sources. In our parameter space, there exists an ultra-small fraction ($\sim 1\%$) IMBH X-ray binaries with initial orbital periods shorter than the bifurcation period, some of which would evolve toward compact X-ray sources. None of these compact IMBH X-ray binaries can appear as ULXs, while they have an opportunity to be visible as low-frequency GW sources \citep{port04b,chen20}.
The influence of the irradiation effect on the evolution of these normal X-ray sources can be ignored owing to the low loss rates of irradiation-driven winds.

\acknowledgments {We are grateful to the referee for her/his valuable comments that have led to the improvement of
the manuscript. We also thank Xiang-Dong Li, and Ying-He Zhao for helpful discussions. This work was partly supported
by the National Natural Science Foundation of China (under grant number 11573016, 11803018, and 11733009), and the CAS "Light of West China" Program (Grants No. 2018-XBQNXZ-B-022). }

\end{document}